\documentclass[12pt,prb,aps,preprint,showkeys,showpacs]{revtex4}
\usepackage[utf8]{inputenc}
\usepackage{amsmath}
\usepackage{amsfonts}
\usepackage{graphicx}
\usepackage[colorlinks=true, citecolor=red, linkcolor=blue, urlcolor=blue]{hyperref}
\begin{document}
\title{Exploration of potential and free energy surfaces of the neutral Be$_4$B$_{8}$ chiral clusters and their stabilities at finite temperatures.}
\author{Carlos Emiliano Buelna-Garc\'ia$^{1,2}$}
\author{C\'esar Castillo-Quevedo$^3$}
\author{Eduardo Robles-Chaparro$^4$}
\author{Tristan Parra-Arellano$^4$}
\author{Jesus Manuel Quiroz-Castillo$^{1}$}
\author{Teresa del Castillo-Castro$^{1}$}
\author{Gerardo Mart\'inez-Guajardo$^{5}$}
\author{Aned de-Leon-Flores$^{4}$}
\author{Gilberto Anzueto-S\'anchez$^{6}$}
\author{Martha Fabiola Martin-del-Campo-Solis$^{3}$}
\author{Ana Maria Mendoza-Wilson$^{7}$}
\author{Alejandro Vasquez-Espinal$^{8}$}
\author{Jose Luis Cabellos$^{9\star}$}\email[email:]{sollebac@gmail.com, josecabellos@uson.mx}
\affiliation{$^1$Departamento de Investigaci\'on en Pol\'imeros y Materiales, Edificio 3G. Universidad de Sonora. Hermosillo, Sonora, M\'exico}
\affiliation{$^2$Organizaci\'on Cient\'ifica y Tecnol\'ogica del Desierto, Hermosillo 83150, Sonora, Mexico}
\affiliation{$^{3}$Departamento de Fundamentos del Conocimiento,
  Centro Universitario del Norte, Universidad de Guadalajara,
  Carretera Federal No. 23, Km. 191, C.P. 46200, Colotl\'an, Jalisco, M\'exico}
\affiliation{$^4$Departamento de Ciencias Qu\'imico Biologicas, Edificio 5A.
  Universidad de Sonora. Hermosillo, Sonora, M\'exico}
\affiliation{$^5$Unidad Acad\'emica de Ciencias Qu\'imicas, \'Area de Ciencias de la Salud, Universidad Aut\'onoma de Zacatecas, Km. 6 carretera Zacatecas-Guadalajara s/n, Ejido La Escondida C. P. 98160, Zacatecas, Zac.}
\affiliation{$^6$ Centro de Investigaciones en \'Optica, A.C., 37150 Le\'on, Guanajuato, M\'exico}
\affiliation{$^7$ Coordinaci\'on de Tecnolog\'ia de Alimentos de Origen Vegetal, CIAD, A.C., Carretera Gustavo Enrique Astiazar\'an Rosas, No. 46, Hermosillo 83304, Sonora, M\'exico}
\affiliation{$^8$ Computational and Theoretical Chemistry Group Departamento de Ciencias Qu\'imicas, Facultad de Ciencias Exactas, Universidad Andres Bello, Republica 498, Santiago, Chile}
\affiliation{$^9$Departamento de Investigaci\'on en F\'isica, Universidad de Sonora, Blvd. Luis Encinas y Rosales S/N, 83000 Hermosillo, Sonora, M\'exico}

\date{\today}
\affiliation{{$^*$corresponding author: sollebac@gmail.com, jose.cabellos@unison.mx}}
\begin{abstract}
  The lowest-energy structure, distribution of isomers, and their
  molecular properties depend significantly on the geometry and temperature.
  The total energy computations under DFT methodology are typically carried out at zero
  temperature; thereby, entropic contributions to total energy are neglected,
  even though functional materials work at finite temperature. In the present study,
  the probability of occurrence of one particular Be$_4$B$_8$ isomer at temperature T is estimated 
   within the framework of quantum statistical
  mechanics and nanothermodynamics. To locate a list of all possible low-energy chiral and achiral structures, an
  exhaustive and efficient exploration of the potential/free energy surface is done by employing
  a multilevel multistep global genetic algorithm search coupled to DFT.  Moreover,
  we discuss the energetic ordering of structures computed at the DFT level against single-point
  energy calculations at the CCSD(T) level of theory. The computed VCD/IR spectrum of
  each isomer is multiplied by their corresponding Boltzmann weight at temperature T;
  then, they are summed together to produce a final Boltzmann weighted spectrum.
  Additionally, we present chemical bonding analysis using the Adaptive Natural Density
  Partitioning method in the chiral putative global minimum. The transition state structures and the
  enantiomer-enantiomer and enantiomer-achiral activation energies as a function of temperature,
  evidence that a change from an endergonic to an exergonic type of reaction occurs at a temperature of 739 K.
\end{abstract}

\maketitle
\section{Introduction}
The potential of boron atom to form stable molecular networks~\cite{Kondo2017,Sudip2} lies in the fact that it has
three valence electrons and four available orbitals, which implies it is electron deficient.
Moreover, it has a small covalent radius of 0.8-1.01~\cite{kabay2015boron,Tahere}~\AA,~ high ionization
energy 344.2 kJ/mol,~\cite{kabay2015boron} and an affinity for oxygen atoms, which is the
basis of borates.~\cite{DeFrancesco,kabay2015boron} Boron electron deficiency gives origin to
vast number of allotropic forms and uncommon geometries~\cite{Wang,Lai-Sheng5,Sudip2} such as
nanotubes,~\cite{Lai-Sheng4,Dong} borospherenes,~\cite{Ying-Jin-Wang} borophene,~\cite{Lai-Sheng5}
cages,~\cite{Dong,Lv} planar,~\cite{Lai-Sheng} quasi planar,~\cite{Lai-Sheng2} rings,~\cite{Bulusu,Dong2019}
chiral,~\cite{Lai-Sheng2,Feng,Qiang,Osvaldo,Chen2,TruongTai} boron-based helix clusters,~\cite{Guo,Feng}
and fluxional boron clusters~\cite{Merino,Sudip2,Jimenez-Halla,Martinez-Guajardo,Moreno,Cervantes-Navarro,Jalife,Guo,C8NA00256H,Gerardo,Schleyer,YU2020112949} that have recently attracted the interest of experimental and theoretical researchers. Since the molecular properties depend greatly on their 
geometry and temperature;~\cite{Baletto,Buelna} boron cluster exhibit
a large number of molecular properties that yield potential applications in
medicine,~\cite{Axtell2018,Fayaz,Hawthorne,Pizzorno2015} molecular motors,~\cite{Jimenez-Halla,Yonggang,Guo}
superhard materials,~\cite{Oganov2009}  hydrogen storage,~\cite{FAKIOGLU20041371}
batteries,~\cite{JIANG201697,LIANG2017152,doi:10.1021/acsami.6b05747,JIANG20181651}
catalysis,~\cite{doi:10.1063/1.4960102} and energy materials~\cite{https://doi.org/10.1002/anie.201911108} among many others.

Particularly, the chirality of nanoclusters has attracted attention
due to their chiroptical properties, potential application in efficient chiral
discrimination,~\cite{Ayuso2019,ayuso2020polarization} nonlinear optics~\cite{C8SC00344K} and chiral materials with
interesting properties,~\cite{doi:10.1021/acs.joc.0c02196,Lai-Sheng2,C7CC08191J}
and of course, not to mention that chiral structures play a decisive role in biological activity.~\cite{Ebeling2018}

Previous theoretical studies joint with experimental photoelectron spectroscopy reported the first pure
boron chiral B$_{30}^-$ structure as the putative global minimum.~\cite{Lai-Sheng2} In these
pair of planar enantiomers, the chirality arises due to the hexagonal hole and its position.
A year later, the lowest energy structures of the  B$_{39}^-$ borospherene were reported as chiral
due to their hexagonal and pentagonal holes.~\cite{Qiang} Similarly, the B$_{44}$ cluster was reported as a
chiral structure due to its nonagonal holes.~\cite{TruongTai}
That is, in these clusters, holes in the structure cause chirality.

Regarding beryllium-doped boron clusters, they exhibit remarkable properties suach as 
fluxionality,~\cite{Guo,YU2020112949,CuiYang,Lai-Sheng5,https://doi.org/10.1002/chem.201203890,C8CP04332A}
aromaticity,~\cite{Guo,D0NJ05961G} and 
characteristics similar to borophene.~\cite{Dongliang}
Furthermore, previous theoretical studies showed that the boron fullerenes B$_{60}$ and B$_{80}$
can be stabilized by surrounding the boron clusters with beryllium atoms,~\cite{GRIBANOVA201944,Gribanova2020}
which effectively compensates for boron electronic deficiency.~\cite{Gribanova2020} These effects make
beryllium-doped boron clusters interesting.

Particularly attractive are the chiral helices Be$_6$B$_{11}^-$, reported by Gou et al.~\cite{Guo},
Yanez et al.,~\cite{Osvaldo} and Garcia-Buelna et al.~\cite{Buelna} as one of the low-lying and fluxional isomers,
and later a chemical bonding and mechanism of formation study of the beryllium-doped boron chiral
cluster Be$_6$B$_{10}^{2-}$ and coaxial triple-layered  Be$_6$B$_{11}^{-}$ sandwich structure were reported.~\cite{Feng,C8CP04332A} In these structures, the chirality arises due to the formation of a boron helix.

However, there are only a few theoretical studies on vibrational circular dichroism (VCD) and infrared spectroscopy (IR) as a function of temperature in beryllium-boron clusters~\cite{C7CP05179D, Buelna}. We emphasize that there are neither theoretical nor experimental studies of VCD / IR spectra in chiral Be$_4$B$_8$ clusters, although VCD/IR spectra give insight into the geometrical structure.~\cite{stephens2012vcd,doi:10.1021/jo9821325,Aamouche,https://doi.org/10.1002/chir.22330} 
  Reciently, Castiglioni et al. reviewed experimental aspects of solid-state circular dichroism,~\cite{https://doi.org/10.1002/chir.20770} and Avil\'es Moreno et al. reported the experimental and theoretical IR/VCD spectra of various compounds.~\cite{https://doi.org/10.1002/cphc.201300503,AVILESMORENO2011767,doi:10.1021/jp801099e,HUET20121261} VCD is differential spectroscopy sensitive to the difference in the absorption for the left and right polarized light.~\cite{stephens2012vcd,Yang2011,https://doi.org/10.1002/chir.22330} It usually is four
times in magnitude smaller than IR absorption~\cite{doi:10.1021/ja00846a061} and yields information on the lowest energy
conformation in solution;~\cite{doi:10.1021/jm401600u} thus, the chiral molecule's absolute configuration can be determined employing the VCD spectra.~\cite{https://doi.org/10.1002/chir.20477}

The IR frequencies are  related to the second derivative of
the potential energy and they are  useful in identifying transition states and computing thermodynamics through  
the vibrational partition function.~\cite{Buelna,Dzib,Kubicki_2019}
Additionally, the structure of neutral boron clusters B$_{11}$, B$_{16}$ and B$_{17}$ was probed by IR.~\cite{Fagiani}. 

The DFT VCD/IR spectra depend on the functional and basis set employed~\cite{Aamouche} and significantly on
the lowest- and the low-energy achiral and chiral structures,
so we need an efficiently sampling of the free energy
surface to know the distribution of isomers at
different temperatures.~\cite{molecules26133953,Buelna,Baletto,Li-Truhlar,Truhlar,Grigoryan} A considerable change
in the isomer distribution and the energetic separation among them is the first notable effect of temperature.~\cite{Buelna,molecules26133953} 
Useful materials work at finite temperatures; in that conditions, Gibbs free energy is
minimized~\cite{10.3389/fchem.2020.00757} and determines the putative global minimum at given temperature,~\cite{Buelna}
whereas, entropy of the atomic cluster is maximized.~\cite{10.3389/fchem.2020.00757} Although in the mid
1960’s, Mermin et al.~\cite{PhysRev.137.A1441} studied
the thermal properties of the inhomogeneous electron gas, most of DFT calculations
are typically performed at zero temperature. Recently, over again, DFT was extended to finite
temperature,~\cite{PhysRevLett.107.163001,GONIS201886,PhysRevB.82.205120}
but nowadays, as far as we know, it is not implemented in any software.
However, molecular dynamics and other simulation tools have been employed to
study atomic clusters at finite temperatures.~\cite{Gerardo,Seitsonen,10.2307/2889981,Jalife,Hill,Fox}

In this study,  based on the Gibbs free energy of each isomer and Boltzmann factors, we computed the probability of
occurrence (Boltzmann weights) of each particular isomer of Be$_4$B$_{8}$ as a function of temperature
using quantum statistical mechanics. The computed VCD/IR spectrum of each isomer is multiplied by their corresponding Boltzmann weight at temperature T; then, they are summed together to produce a final Boltzmann weighted spectrum.
In the mid 1980, P. J. Stephens with co-workers implemented the atomic axial tensors
in Gaussian 80 code that allows them to compute the VCD spectrum of propylene oxide and compare with the experimental
spectrum~\cite{Stephens2008}
Later, Nafie and Stephens employed the Boltzmann weights scheme, they computed the VCD spectrum for each isomer, and the total resulting spectra were averaged and weighted by Boltzmann
factors.~\cite{nafie2011vibrational,stephens2012vcd,doi:10.1146/annurev.physchem.48.1.357,doi:10.1366/11-06321}
Recently these factors were used in other previous works.~\cite{Li-Truhlar,Buelna,Truhlar,PhysRevLett.107.163001,Grigoryan,calvo}.

To achieve the mentioned above, we
located all low-energy structures on the potential and free energy surfaces of the Be$_4$B$_{8}$ cluster
with a genetic algorithm coupled to DFT and computed the Boltzmann
weights in temperatures ranging from 20 to 1900 K. We also located the
solid-solid transformation point at 739 K, where chiral and achiral structures coexist, and computed
the energy barrier (E$_a$) for temperatures ranging from 20 to 1900 K for transformation
of enantiomers (plus, $\mathcal{P}$) to an achiral structure. We locate the T$_{ee}$ point is defined here as the temperature where the reaction change from endergonic to exergonic. Moreover, the energy of enantiomerization was
computed between $\mathcal{P}$ and minus ($\mathcal{M}$) enantiomers, and we defined the T$_{bb}$ point in scale temperature where the energy barrier of two possible reaction mechanisms is equal to each other, which implies the velocity of the reaction is equal for both mechanisms.
We investigated how the symmetry point group affects the Gibbs free energy. 
Our results show that the chirality on
Be$_4$B$_{8}$ arises from the  Be-Be dimers capping the boron ring  and also of the distorted boron ring,
thus, the lowest energy chiral structure is favored by  the interaction between beryllium and the boron framework.
The high energy of enantiomerization of the Be$_4$B$_8$ cluster in temperatures ranging from 20 to 1900 K
suggests that it is a good candidate for applications; only one of the enantiomers shows the desired effect.
The computed  enthalpy of formation ($\Delta$H) between chiral and achiral structure at 739 K
show that there is a change from endo to exothermic reaction.
Our results indicate that the Boltzmann  weighted VCD spectrum is zero in all
range of temperature, wheras, the Boltzmann IR weighted spectrum is strongly dominated by the lowest
energy pair of enantiomers. he remainder of the manuscript is organized as follows: 
Section 2 gives the computational details and a brief overview of the theory and the algorithms used.
The results and discussion are presented in section 3. We discuss the effect of the symmetry in the
energetic ordering and clarify the origin of the 0.41 kcal/mol difference energy between two
structures with symmetries C$_2$ and C$_1$ that appear when we compute the Gibbs free energy.
A comparison among energies computed at a single point CCSDT  against DFT levels of theory and the
$\mathcal{T}_1$ diagnostic is presented. We do the chemical bonding analysis by employing the AdNDP scheme
to $\mathcal{P}$ minimum energy structure.
The interconversion energy barrier  between the  $\mathcal{P}$
and  $\mathcal{M}$ enantiomers and between an achiral structure and $\mathcal{P}$  enantiomer are discussed in
terms of temperature. IR spectra are analyzed as a function of temperature. Conclusions are
given in Section 4.
\section{Theoretical Methods} 
\subsection{Global Minimum Search and Computational Details}
First of all, for theoretical studies of an atomic cluster, the first step is
locating the putative global minimum and all the closet low-energy structures on its potential/free energy surface, since the measured molecular properties are statistical averages over the ensemble of conformations.~\cite{Teague,Buelna} We must keep in mind that experimental atomic molecular studies are conducted in non-zero temperatures while theoretical studies based on density functional theory computations are performed at zero temperature.~\cite{computation4020016} If the temperature increases, Gibbs free energy determines the lowest-energy structure,~\cite{Buelna} and the entropy of the atomic cluster
is maximized,~\cite{10.3389/fchem.2020.00757} while at temperature zero, the enthalpy determines the putative global minimum. Secondly, the search of the global minimum in atomic clusters is a complicated task due to the degrees of liberty increase as a function of the number of atoms. Consequently, the number of possible combinations grows exponentially,
leading to a combinatorial explosion problem.~\cite{Morris} Thirdly, a
systematic and
exhaustive exploration of the multidimensional potential/free energy surface, avoiding an incomplete sampling of the configuration space is needed.~\cite{Truhlar,Li-Truhlar,Buelna}  Finally,  it is important  to take into account all low-energy structures with all low-symmetries due to the influence  of  the point group symmetry of the molecule on the Gibbs free energy through rotational entropy, which is a function of the symmetry number that appears in the rotational entropy denominator.  This could affect the relative populations leading to miscalculation of molecular properties when they are computed employing weighted Boltzmann factors.~\cite{Buelna}  Despite the difficulties mentioned above, several algorithms to explore the potential/free energy surface coupled to an any electronic structure package have been successfully employed  so far, such as \emph{AIRSS} approach,~\cite{Pickard_2011} simulated annealing,~\cite{kirkpatrick, metropolis, xiang, yang, vlachos, granville} kick methodology~\cite{Sudip,Cui,Vargas-Caamal2,Vargas-Caamal,Cui2,Vargas-Caamal2015,Florez,Ravell,Hadad,Saunders,Saunders2,Grande-Aztatzi,https://doi.org/10.1002/slct.201600525} and  genetic algorithms~\cite{Guo,Dong,Mondal,https://doi.org/10.1002/chem.201702413,C8CP01009A,Ravell,Grande-Aztatzi,Kessler,Alexandrova,Buelna,C7CP01328K} among others. Our computational procedure to
explore the potential/free energy surface  of the neutral Be$_4$B$_8$ cluster employs a hybrid genetic algorithm
implemented in ~\emph{GALGOSON} code.~\cite{Buelna,molecules26133953}
This methodology based on previous works~\cite{kcabellos,Hernandez,Guo,Grande-Aztatzi} consists of a multi-step approach (cascade) to efficiently sample the potential/free energy surface coupled to the \emph{Gaussian}~\cite{gauss} code.
Our multi-step strategy
employs  more accurate levels of theory applied to each step to arrive at the most stable lowest-lying isomers.
In the first step of our methodology, the code builds an initial random population of planar and 3D structures
(two hundred structures per atom of the Be$_4$B$_8$ cluster) employing a strategy used in previous
work.~\cite{Grande-Aztatzi,Guo,Buelna,Hernandez,Kessler,Mondal,Castro,Hadad}
Then, structures are optimized at the PBE0\cite{Adamo}/LANL2DZ~\cite{schaefer1977methods} level of theory with
Gaussian 09 code.
Here we consider that the overall global search methodology might be sped up using a smaller basis set, in early stage.
The PBE0 functional has shown good performance in pure and doped
boron clusters,~\cite{Zhao-Jijun,Jeng-Da,Guo,Buelna,Osvaldo}, moreover, energetic analisis of anionic Be$_6$B$_{11}$ cluster in several previous works were done employing PBE0 functional~\cite{Guo,Buelna,Osvaldo}
Additionally, relative energies at the PBE0 level are similar  to those the  CCSD(T) level in  B$_9^{-}$ boron cluster.~\cite{Li-Li}
As a second step, with the aim of increaasing the performance, and  deciding if
a structure should be further relaxed;~\emph{GALGOSON}
makes a cluster shape comparison among them by a superposition method~\cite{Rush};
All  isomers that lie below 20 $\cdot$mol$^{-1}$ on the energy scale 
are permitted to relax, and they compose the first cluster population of the genetic algorithm.
The optimization in this step is done at the PBE0~\cite{Adamo}/def2TZVP~\cite{Weigend,Ansgar} level of theory,
including Grimme's dispersion (GD3) effects~\cite{Grimme} as implemented in Gaussian 09 code.
The criterion to stop the algorithm is that the lowest energy structure persists for five generations.
In the third step, structures lying within 20 kcal$\cdot$mol$^{-1}$ found in the previous step are symmetrized at the
low-symmetry point group. Those structrues are the initial population for the genetic algorithm at  the PBE0-GD3/def2-TZVP
level, taking into account the zero-point energy (ZPE) corrections. In total at  this point and in all previous stages, about ~2800 relaxations to a local-energy minimum were performed. Additionally, we make sure that the lowest vibrational mode of each isomer is positive to identify a valid energy minimum and discard transition states.
In the final step,  single point (SP) computations  for the low-energy structures lying below 8 kcal$\cdot$mol$^{-1}$
carried out 
at CCSD(T)/def2-TZVP//PBE0-D3/def2-TZVP level of theory
as it is implemented in Gaussian 09 code.~\cite{gauss} Moreover, SP were computed employing the domain-based local pair natural orbital coupled-cluster theory (DLPNO-CCSD(T)), with and without taking into account the ZPE correction energy. Furthermore, 
to determine if the energy evaluation scheme based on a single reference method of the Be$_4$B$_8$ cluster, contains a substantial multireference character. We computed  the $\mathcal{T}_1$ diagnostic.  Our results confirm that the computed $\mathcal{T}_1$ diagnostic values are below the recommended threshold of 0.02~\cite{Lee,Hernandez}  for all low-energy isomers and ensure that DFT energies of Be$_4$B$_8$ do not contain a large multireference character. Hernandez et al.~\cite{Hernandez} found similar values for $\mathcal{T}_1$ descriptor in doped Boron clusters. The $\mathcal{T}_1$ diagnostic and the SP calculations at DLPNO-CCSD(T) level were performed using ORCA program suite,~\cite{Liakos} with \emph{TightPNO} settings.~\cite{Liakos2} Chemical bonding was examined using the Adaptive Natural Density Partitioning (AdNDP) method.~\cite{B804083D} AdNDP analyses the first-order reduced density matrix and recovers Lewis bonding (1c-2e or 2c-2 e, i.e., lone pairs or two-center two-electron bonds) and delocalized bonding elements associated with the concept of electron delocalization. 

\subsection{Thermochemistry Properties}
The molecular partition function contains all thermodynamic information in a similar way that the wavefunction contains all the information about the system.~\cite{Buelna,Dzib} which implies that all thermodynamic properties of an ensemble of molecules can be derived from this function.  
Previous theoretical studies used the partition function to compute temperature-dependent entropic contibutions~\cite{Brehm} on [Fe(pmea)(NCS)2] complex, infrared spectroscopy on anionic Be$_6$B$_{11}$ cluser,~\cite{Buelna} and rate constant.~\cite{Dzib} In this study, the thermodynamic functions dependent on temperature are computed employing the partition function Q shown in Equation~\ref{partition}, and computed under the rigid rotor, harmonic oscillator, Born-Oppenheimer, ideal gas, and a particle-in-a-box approximations.
\begin{equation}
\displaystyle
Q(T)=\sum_{i}g_i~e^{{-\Delta{E_i}}/{K_BT}}
\label{partition}
\end{equation}
In Eq.~\ref{partition}, $g_i$ is the degeneracy factor, $k_{\textup{B}}$ is the Boltzmann constant, $T$ is the temperature, and ${-\Delta{E_i}}$ is the total energy of a cluster.~\cite{Buelna,Dzib,mcquarrie1975statistical} We employ equations~\ref{e1} to~\ref{e4} to compute the internal energy (U), enthalpy (H), and Gibbs energy (G) of the Be$_4$B$_8$ cluster at temperature T. The equations 1 to 4  and the equations to compute  entropy contributions (S) are those employed in a previous work~\cite{Buelna,molecules26133953,Dzib} and any standard  thermodynamics textbook.~\cite{mcquarrie1975statistical,hill1986introduction}
\begin{equation}
\displaystyle
\mathcal{U}_0=\mathcal{E}_0+ZPE\\
\label{e1}
\end{equation}
%%%
\begin{equation}
\displaystyle
U_T=\mathcal{U}_0+(E_{ROT}+E_{TRANS}+E_{vib}+E_{elect})\\
\label{e2}
\end{equation}
%%%
\begin{equation}
\displaystyle
H=U_T+RT\\
\label{e3}
\end{equation}
%%%
\begin{equation}
\displaystyle
{G}=H-TS\\
\label{e4}
\end{equation}
n Equations above, ZPE is the zero-point energy correction  $\mathcal{E}_0$ is the electronic energy,
and $E_{ROT}+E_{TRANS}+E_{VIB}$ are the contributions to energy due to
translation, rotation, and vibration as function of temperature, respectivly.
In order to compute the probability of occurrence of one particular cluster
in an ensemble of Be$_4$B$_8$ clusters (Boltzmann ensemble at thermal equilibrium) 
as a function of temperature, we employ the probability of occurrence ~\cite{Buelna,Truhlar,Bhattacharya,Bhumla,Shortle,MENDOZAWILSON2020112912,Dzib,Schebarchov,Goldsmith,Grigoryan} %,Zhen}
given in Equation~\ref{boltzman1} 
\begin{equation}
\centering 
\displaystyle
P_i(T)=\frac{e^{-\beta \Delta G^{k}}}{\sum e^{-\beta \Delta G^{k}}}\label{boltzman1},
\end{equation}
where $\beta=1/k_{\textup{B}}T$, and $k_{\textup{B}}$ is the Boltzmann constant, $T$ is the temperature, and $\Delta G^{k}$ is the Gibbs free energy of the $k^{th}$ isomer,  Any molecular properties observed  are statistical averages over a Boltzmann ensemble of clusters, then for an ensemble of clusters, any property can
be computed as an average of all possible conformations.~\cite{Buelna,MENDOZAWILSON2020112912}
Equation~\ref{boltzman1} is restricted so that  the sum of all
probabilities of occurence, at fixed temperature T, $P_i(T)$ must be equal to 1 given, according
to Equation~\ref{bol2}
\begin{equation}
\centering 
\displaystyle
\sum_i P_i(T)=1\label{bol2},
\end{equation}
In this study, the Boltzmann  weighted VCD/IR  spectrum (VCD/IR$_{Bolt}$) at temperature T is given by Equation~\ref{vcd}
\begin{equation}
\displaystyle
VCD/IR_{Bolt}=\sum_i^{n}VCD_{i}/IR_{i}\times P_i(T)
\label{vcd}
\end{equation}
Where $n$ is the total number of cluster in the ensemble, VCD/IR$_{i}$ is the VCD/IR of the $i^{th}$  isomer at temperature T=0, and  P$_i$(T) is the probability of the $i$ isomer given by Equation~\ref{boltzman1}. The sum only runs over all achiral, Plus and Minus isomers. For achiral structures the VCD is equal to zero and there is no contribution to  $VCD_{Bolt}$. Here, we point out that is important to take into account the
achirial structures due to the probability of a particular chiral cluster changing as consequence of 
the  $VCD_{Bolt}$. in spite of the  VCD for achiral structures being zero.
For the computation of relative populations and VCD/IR$_{Bolt}$ spectra, 
we used the Boltzmann-Optics-Full-Ader code (\emph{BOFA}) that is part of
spectroscopy capabilities of \emph{GALGOSON} code.~\cite{Buelna}
%%%%%
%%%%%
%%%%%
\section{Results and Discussion}
\subsection{The lowest energy structures and energetics}
\begin{figure}[ht!]
\begin{center}  
  \includegraphics[scale=0.80]{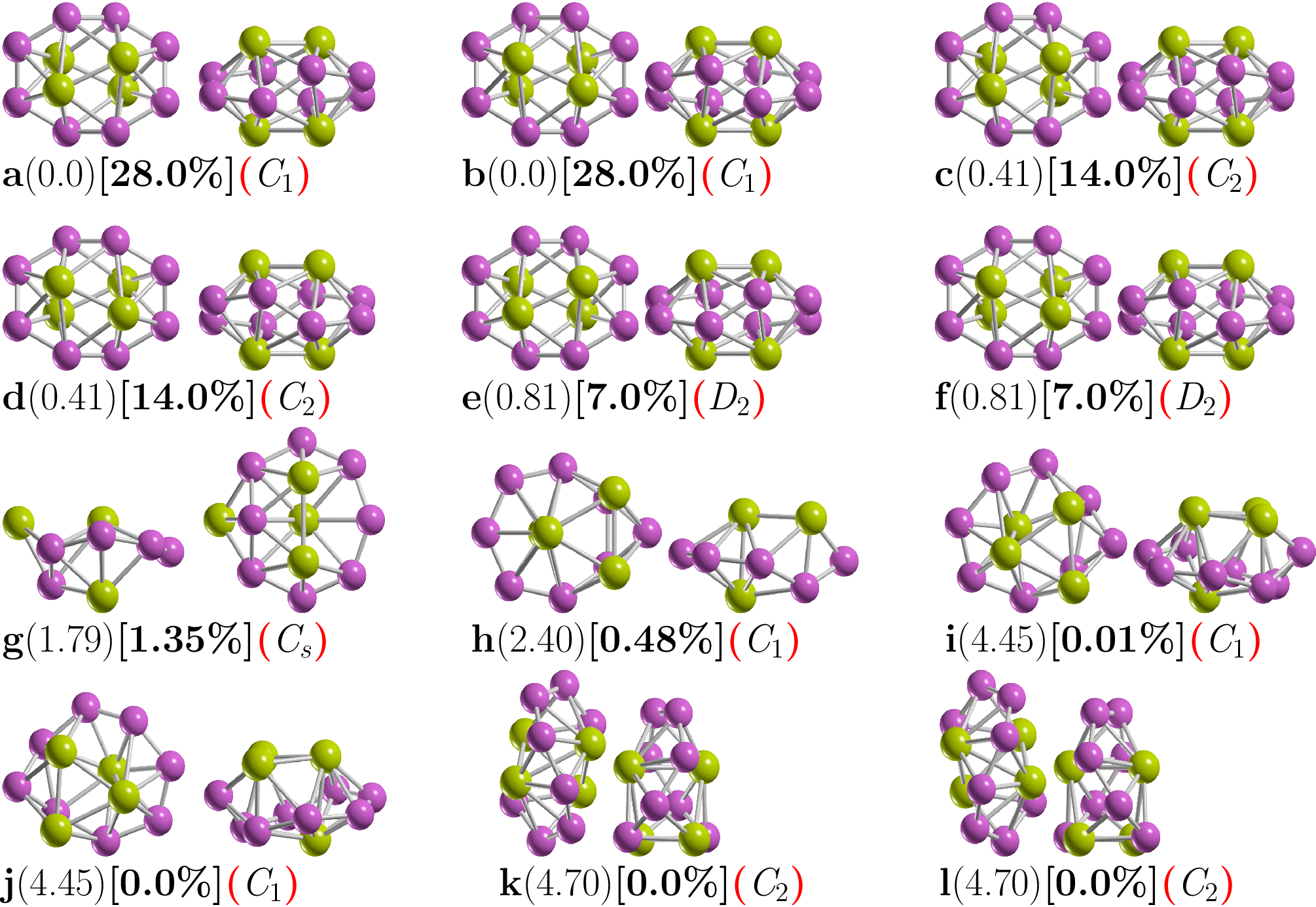}
\caption{(Color online) Optimized geometries of neutral Be$_{4}$B$_{8}$ cluster at PBE0-GD3/def2TZVP level of theory with ZPE correction energy. They are shown in two orientations, front, and rotated 90 degrees up to paper plane.
  The lower-case letter is the isomer label, relative Gibbs free energies in kcal/mol (in round parenthesis) at 298.15 K,
  the relative population [in square parenthesis], and group symmetry point (in red round parenthesis). The structures with label (a,b), (c,d), (e,f), (i,j) and (k,l) are chiral. The purple- and yellow-colored spheres represent the boron and beryllium atoms, respectively.~\cite{Buelna} Atomic Cartesian coordiantes of these isomers are provided in Supporting Material.}
\label{geometry_gibbs}
\end{center}    
\end{figure}

\begin{figure}[ht!]
\begin{center}  
\includegraphics[scale=0.35]{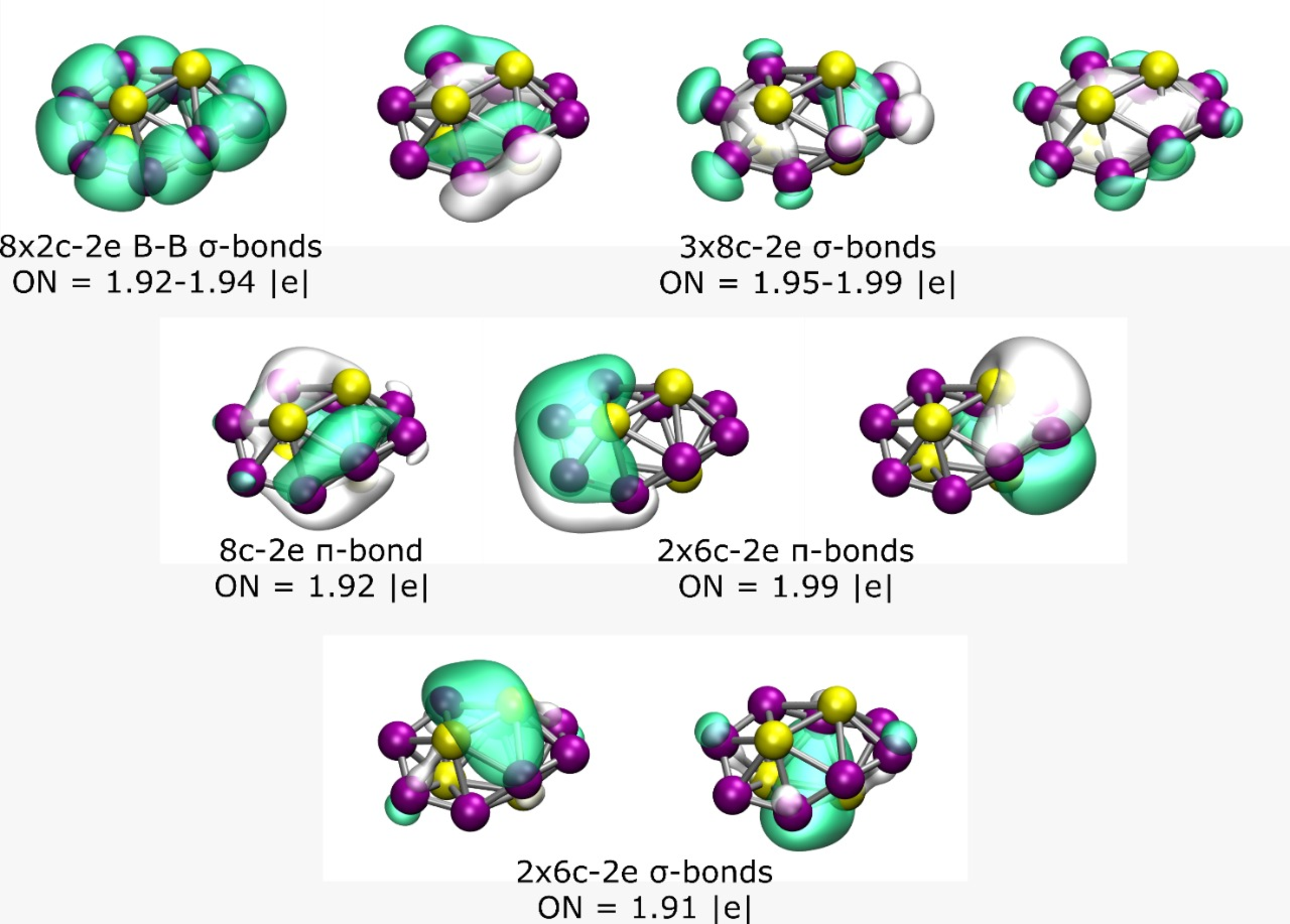}
\caption{(Color online) Results of the AdNDP analysis of the lowest-energy chiral isomer of the Be$_4$B$_8$ system.} 
\label{adn}
\end{center}
\end{figure}

Figure~\ref{geometry_gibbs} shows the low-energy configurations of Be$_4$B$_{8}$ clusters optimized
at PBE0-GD3/def2-TZVP level of theory taking into account ZPE energy correction.
The optimized average B-B bond length of the putative chiral global minimum is 1.5867~\AA,~in good agreement with an experimental bond length of 1.57-1.59~\AA.~\cite{Moezzi,Zhou} and also within agreement with others previous DFT calculations.~\cite{Buelna} The most recurring motif within the lower energy isomers of B$_8$Be$_4$ is a sandwich structure,(SSh) in which
the boron atoms form a hollow distorted ellipsoid ring with each of the Be-Be dimers capping the top and bottom 
with C$_1$ point group symmetry.
Isomers $a$ and $b$ and also listed as $i_1$ and $i_2$ in Table~\ref{energia}, are enantiomers differing in the orientation of the Be-Be dimers with respect the boron skeleton. Based on the B-B bond length evolution along the intrinsic reaction coordinate (IRC) (See movie in Supporting Material)  between Plus-enantiomer-Minus-enantiomers, and displayed in Figure~\ref{evo} appendix~\ref{BeBe} the shortest B-B bond length is located at the transition state structure. In contrast, the largest  B-B bond length is located in the reactant and product points.  On the other hand, appendix Figure~\ref{evo} appendix~\ref{BeBe} shows the distance evolution between (Be-Be)-(Be-Be) dimers; one can see the largest distance between dimers is located at the transition state, whereas the shortest distance is located at the product and reactants points.   From the above mentioned, the B-B interaction does not favor the formation of the lowest energy enantiomers structures; meanwhile, the  Be-Be interaction promotes the lowest energy structure to be chiral. Here, we infer that the Be-B interaction also favors the chiral lowest energy structures. The Be-Be bond length for the six lowest energy enantiomers  is 1.9874, 1.9876, and 1.9881~\AA~ for symmetries C$_1$, C$_2$, and D$_2$, respectively, in good agreement with the bond length of the Be-Be in Be$_2$B$_8$ cluster 1.910~\AA.~\cite{CuiYang}  To gain more insight into the chemical bonding, an AdNDP analysis of the lowest energy isomer was performed (Figure~\ref{adn}). The AdNDP analysis for this chiral structure revealed the presence of eight 2c-2e B-B $\sigma$-bonds with an occupation number (ON) between 1.92 and 1.94 |e| and three delocalized $\sigma$-bonds throughout the B$_8$ ring with an ON between 1.95 and 1.99 |e|. Additionally, three distorted $\pi$-bonds (due to the non-planarity of the structure), one of which is delocalized over all eight boron atoms and the other two involving four boron and two beryllium atoms (one from the top and one from the bottom). Finally, the bonding pattern is completed by two 6c-2e $\sigma$-bonds with main contribution coming from the interaction between the two Be atoms from the top and bottom, respectively.
The isomers with symmetry C$_1$ are the most energetically favorable, with 28{\%} each of the
Boltzmann population at 298.15 K. An exhaustive and systematic exploration of the potential
energy surface considering triplet states revealed that the lowest triplet ground state lays
13.7 kcal/mol above the singlet putative chiral global minimum single ground state,
which is too far away energetically to be considered.
Next, low-energy SSh-isomers labeled $i_3$ and $i_4$ in Table~\ref{energia}, and depicted in
Figure~\ref{geometry_gibbs}(3,4) lies just 0.41 kcal/mol above the putative global minimum,
it is a similar-SSh structure than the putative global minimum, only with C$_2$ point group symmetry,
followed by slightly higher energy and similar-SSh structure located just 0.81 kcal/mol
above the putative minimal structure with D$_2$ point group symmetry.
We point out that the unique differences among these chiral structures are the different symmetry point groups.
The most energetically favorable is the non symmetry (C$_1$) cluster; moreover, these six structures contribute to
98{\%} to the relative population at 298.15 K.
\begin{table}[!ht]\centering
  \caption{Single point relative energy calculations of the low-energy structures at differents levels of theory. Coupled Cluster Single-Double and perturbative Triple, (CCSD(T)), CCSD(T) with zero point energy ($\mathcal{E}_{\mathrm{ZPE}}$), (CCSDT{\normalsize{$+\mathcal{E}_{\mathrm{ZPE}}$})},  CCSD(T) employing  the domain-based local pair natural orbital coupled-cluster theory (DLPNO-CCSD(T)), with TightPNO setting,
   with zero point energy (DLPNO-CCSD(T){\normalsize{$+\mathcal{E}_{\mathrm{ZPE}}$})}), Gibbs free energy ($\Delta G$) at 298.15 K, Electronic energy with $\mathrm{ZPE}$ {(\normalsize{$\mathcal{E}_0+\mathcal{E}_{\mathrm{ZPE}}$)}}, Electronic energy {(\normalsize{$\mathcal{E}_0$)}}, point group symmetry, and \normalsize{$\mathcal{T}_1$ Diagnostic.} All relative energies are given in kcal$\cdot$mol$^{-1}$ }
\label{energia}
 \begin{tabular}{@{\extracolsep{0.2pt}} ll l l lll lll l }
 \\[-1.8ex]\hline 
 \hline \\[-1.8ex] 
  &   &  \multicolumn{9}{c}{\normalsize{Isomers}}\\
\cline{2-11}\\ [-1.8ex] 
\multicolumn{1}{c} {{{ \normalsize{ Level }}}}  & \multicolumn{1}{l}{$i_1$} & \multicolumn{1}{c}{$i_2$} & \multicolumn{1}{c}{$i_3$ } & \multicolumn{1}{c}{$i_4$}& \multicolumn{1}{c}{$i_5$}& \multicolumn{1}{c}{$i_6$}  &\multicolumn{1}{c}{$i_7$} &\multicolumn{1}{c}{$i_8$}&\multicolumn{1}{c}{$i_9$} & {$i_{10}$}  \\
\hline \\[-1.8ex]
$\Delta G$                                          & 0.0  &  0.0  & 0.41  & 0.41 & 0.81&  0.81   & 1.79 & 2.40 & 4.45 & 4.45 \\
CCSDT                                               & 0.0  &  0.0  &  0.0  &  0.0  & 0.0   & 0.0  & 3.61 & 3.38 & 5.38 & 5.38 \\
CCSDT\normalsize{$+\mathcal{E}_{\mathrm{ZPE}}$}        & 0.0  &  0.0  &  0.0  &  0.0  & 0.0   & 0.0  & 2.71 & 2.51 & 4.51 & 4.51 \\
DLPNO-CCSD(T)                                       & 0.0  &  0.0  &  0.0  &  0.0  & 0.0   & 0.0  & 0.75 & 1.37 &  5.0   & 5.0 \\
DLPNO-CCSD(T)\normalsize{$+\mathcal{E}_{\mathrm{ZPE}}$}& 0.0  &  0.0  &  0.0  &  0.0  & 0.0   & 0.0  & -0.20 & 0.50 & 4.10   & 4.10\\
\normalsize{$\mathcal{E}_0+\mathcal{E}_{\mathrm{ZPE}}$}& 0.0  &  0.0  &  0.0   & 0.0  & 0.0   & 0.0  & 2.38 & 2.80 & 5.03 & 5.03\\
\normalsize{$\mathcal{E}_0$}                        & 0.0  &  0.0  &  0.0   & 0.0  & 0.0   & 0.0  & 3.28 & 3.68  & 5.90 & 3.28\\
{Point group symmetry}                              & {\normalsize{\emph{C$_{1}$}}}  &  {\normalsize{\emph{C$_{1}$}}} &  {\normalsize{\emph{C$_{2}$}}}   & {\normalsize{\emph{C$_{2}$}}}
& {\normalsize{\emph{D$_{2}$}}}         &  {\normalsize{\emph{D$_{2}$}}} & {\normalsize{\emph{C$_{s}$}}}   & {\normalsize{\emph{C$_{1}$}}} & {\normalsize{\emph{C$_{1}$}}}& {\normalsize{\emph{C$_{1}$}}}   \\
\normalsize{$\mathcal{T}_1$ Diagnostic}             & 0.019  &  0.019  &  0.019   & 0.019  & 0.019   & 0.019  & 0.016 & 0.015  & 0.016 & 0.016\\
\hline \\[-1.8ex]
\end{tabular}
\end{table}
The next higher energy structure, labeled as $i_7$ in Table~\ref{energia} and depicted in
Figure~\ref{geometry_gibbs}(g), is located at 1.79 kcal/mol above the putative minimum global at 298.15 K,
with  symmetry C$_s$. It is also a sandwich structure  formed by a distorted circular ring in which one of the
Be-Be dimers is capping in the center of the ring, and the other one is located on one face of the
boron circular ring. This structure is achiral, and its probability of occurrence is 1.35{\%} at 298.15 K.
Next achiral isomer lies 2.40 kcal/mol above the putative minimum global with C$_1$ symmetry, labeled as $i_8$
in Table~\ref{energia} and depicted in Figure~\ref{geometry_gibbs}(8). It is also a sandwich-type structure
formed by a distorted circular  boron ring with three boron atoms capping one side of the ring and the other
Be atom  capping the other. The probability of occurrence of this isomer at 298.15 K is
just 0.48{\%} and its contribution to chiroptical spectroscopies is negligible. The next two chiral structures
lies 4.45 kcal/mol above the putative  global minimum with C$_1$ symmetries,  labeled as $i_9$ and $i_{10}$ in Table~\ref{energia} and depicted in Figure~\ref{geometry_gibbs}(i,j). They are sandwich-type structures formed by a non-planar distorted circular boron ring with three Be atoms capping one side of the boron ring and the other Be atom is located on the other face, and in the center of the distorted boron ring; Its Boltzmann probability of occurrence is zero at 298.15 K, so as a consequence, at this temperature, its contributions to any chiroptical spectroscopies are negligible. The following chiral higher energy structure, with C$_2$ point group symmetry, lies 4.70 kcal/mol energy above the putative global minimum. It is a chiral helix type structure depicted in Figure~\ref{geometry_gibbs}(12,13); it has four Be atoms located in the center of the boron spiral, this helix structure is similar to those found by previous theoretical works~\cite{Guo,Osvaldo,Buelna} and its probability of occurrence is negligible at room temperature. To gain insight into the energy hierarchy of isomers and validate our DFT calculations,  relative energies were computed at different levels of theory, and differences between them are shown in Table~\ref{energia}.
Energy computed at different methods yield
different energies due mainly to the functional and basis-set employed,~\cite{Puente,Buelna}, so the energetic ordering change; consequently, the probability of occurrence and the molecular properties will change.
The first line of Table~\ref{energia} shows the relative Gibbs free energy computed at PBE0-GD3/def2-TZVP and room temperature. The small relative Gibbs free energies (0.41, and 0.81 kcal/mol) differences among the six enantiomer structures $i_1$ to $i_6$ in Table~\ref{energia} are caused by the rotational entropy being a function of the symmetry number that in turn depends on the point group symmetry.  An increase/decrease in the value of rotational entropy changes the Gibbs free energy.  The Gibbs free energy computed with and without symmetry will differ by a factor RT$\ln$($\sigma$). Here, R is the universal gas constant, T, the temperature, and $\sigma$ is the symmetry number.
\begin{figure}[ht!]
\begin{center}  
\includegraphics[scale=0.80]{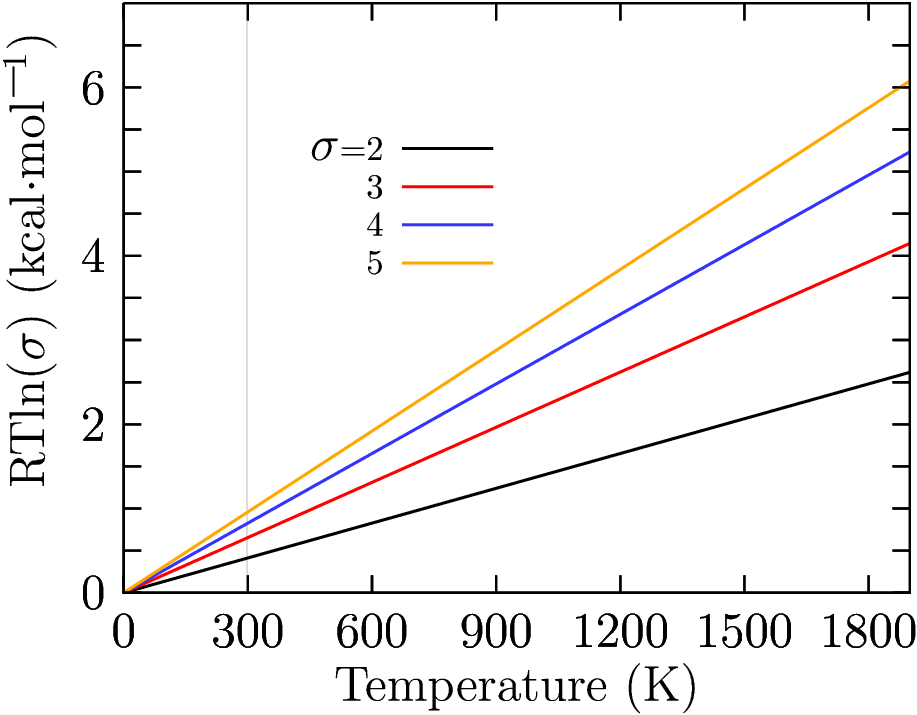}
\caption{(Color online) The difference of the rotational entropy computed with and without symmetries is given by a factor of RT$\ln(\sigma)$  in kcal/mol; in this factor,  R is the universal gas constant, T, the temperature, and $\sigma$ is the symmetry number (The factor is similar to the enantioselectivity~\cite{Peter}). These values, at 298.15 K,  are in good agreement with the values shown in the first line of Table~\ref{energia} (0.41and 0.81 kcal/mol).} 
\label{rot2}
\end{center}
\end{figure}
Figure~\ref{rot2} shows the factor RT$\ln$($\sigma$) for 
temperatures ranging from 0 to 1900 K and for different symmetry number values ($\sigma$=2,3,4,5).
A closer analysis of Figure~\ref{rot2}, shows that at room temperature  RT$\ln(\sigma)$=0.41 kcal/mol with $\sigma$=2, and RT$\ln(\sigma)$=0.81 kcal/mol with $\sigma$=4, in agreement with the values shown in the first line of Table~\ref{energia}.   As the temperature increases, the energy differences between the factors RT$\ln(\sigma)$ become larger.
These small relative Gibbs free energies are responsible for differents values of probability of occurrence at low temperatures for the similar isomers with different point group symmetry. This strongly suggests that there must be atomic clusters with low and high symmetries in the Boltzmann ensemble to compute the molecular properties correctly.  The second line in Table~\ref{energia} shows single point (SP) relative energies computed at the CCSD(T)\cite{Pople}, the energetic ordering of isomers listed in the first line of Table~\ref{energia} follows almost the trend of energetic ordering at SP CCSD(T) level, notice that just the achiral isomers label $i_7$ to $i_8$ in Table~\ref{energia} are interchanged in energetic ordering. 
The third line  Table~\ref{energia} shows single point relative energies computed at the CCSD(T)\cite{Pople}/def2-TZVP//PBE0-GD3/def2-TZVP; the energetic ordering is similar to pure CCSD(T) energy.  DLPNO-CCSD(T)) realtive energies, with and without ZPE correction, are shown in lines four and five of Table~\ref{energia}, the first follows the trend of pure CCSD(T) energy, and the second, the ZPE value, interchange the isomers, label $i_7$ in Table~\ref{energia}, to be the putative global minimum. Here we can say that the ZPE energy inclusion is essential in distributing isomers and molecular properties.
The sixth and seventh lines of Table~\ref{energia} show the electronic energy with and without ZPE correction, and both of them
follow the trend of the Gibbs free energy given in line number one. Line number 8 in Table~\ref{energia} shows the point group
symmetry for each isomer. The T$_1$ diagnostic for each isomer is shown inline nine of Table~\ref{energia}, all of them are lower than the recommended value 0.02~\cite{Pople, Hernandez} so the systems are appropriately characterized.

\subsection{Structures and Stability at Finite Temperature.}
\begin{figure}[ht!]
\begin{center}  
\includegraphics[scale=0.85]{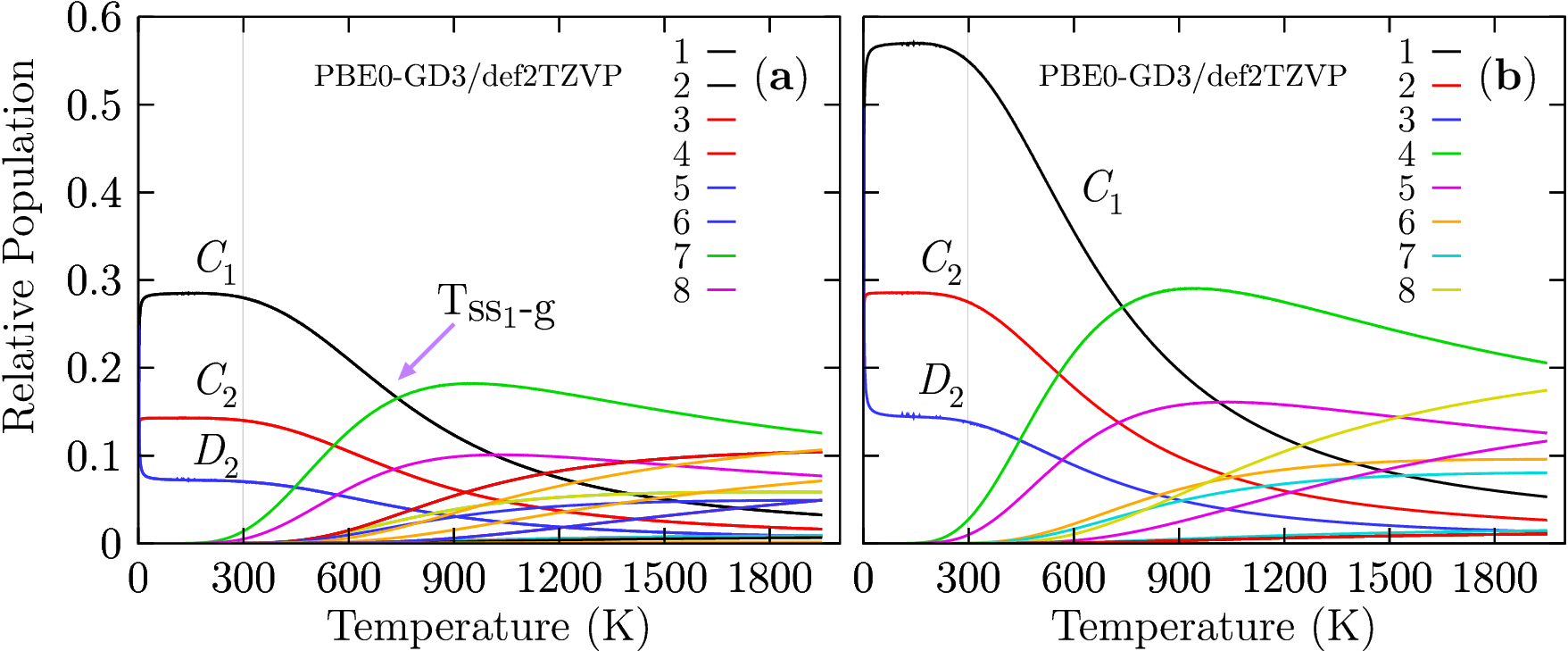}
\caption{(Color online) The left panel (a) shows %the relative population
  the probability of occurrence 
  for temperatures ranging from 20 to 1900 K at the PBE0-GD3/def2-TZVP level of theory considering that the Boltzmann ensemble is composed of achirial structures and an equal
  mixture of $\mathcal{P}$, and $\mathcal{M}$ enantiomers which implies, a Boltzmann racemic ensemble.
  The right panel (b) shows the probability of occurrence for temperatures ranging from 20 to 1900 K at the
  PBE0-GD3/def2-TZVP level of theory taking into account only the achirial structures and $\mathcal{M}$ enantiomers,
  which implies Boltzmann pure ensemble of only one enantiomer.
  percent enantiomeric excess is 100{\%}. (Boltzmann pure ensamble of only one enantiomer)
  In panel (a) the transition solid-solid point ({\large{T$_{\textrm{ss$_1$-g}}$}}) is located at 739 K with 16.6{\%} of probability, while in panel (b) the {\large{T$_{\textrm{ss$_1$-g}}$}} is located at 739 K with 27{\%} probability. The lowest-symmetry  \emph{C$_1$} is strongly dominating at temperatures from 20 to 739 K due to rotational entropy that is a function of the point group symmetry.} 
\label{popu}
\end{center}
\end{figure}
As we mentioned earlier, the determination of the structure is the first step to study any property of a material. Moreover, we have to consider that an observed molecular property in a Boltzmann ensemble is a weighted sum of all individual contributions of each isomer that form the ensemble. At temperature 0 K, the electronic energy plus zero-point energy determine the putative global minimum and all nearby low-energy structures (PGMLES), whereas, at temperatures larger than 0 K, the Gibbs free energy defines the PGMLES. Figure~\ref{popu} shows the probability of occurrence for each particular chiral and achiral Be$_4$B$_8$ isomers for temperatures ranging from 20 to 1900 K. In panel (a) the probability of occurrence is shown, taking into account the $\mathcal{M}$,$\mathcal{P}$, and achiral structures, which implies Boltzmann racemic ensemble 
(the percent enantiomeric excess is zero (Boltzmann racemic ensemble).
whereas panel (b) shows the probability of occurrence only taking into account just the $\mathcal{M}$ handled and achiral structures, which implies
that the percent enantiomeric excess is 100{\%}. then the
ensemble is a Boltzmann pure ensemble of only one type of enantiomer.
There is a significant difference in the probability of occurrence between the two panels. In panel (a), we consider the $\mathcal{P}$ and $\mathcal{M}$ structures, and both structures possess the same probability of occurrence in all ranges of temperature. All the  probabilities of occurrence (chiral) 
shown in panel (b) are approximately two times the probability of occurrence (chiral)   shown in panel (a). A closer examination of the panel (a) shown that in the temperature ranging from 20 to 300 K, all molecular properties are dominated by chiral structure depicted in Figure~\ref{geometry_gibbs}(a,b) becuase its probability of occurrence is almost constant. We point out that in this range of temperature, the C$_1$, C$_2$ and D$_2$ symmetries strongly dominate with different probabilities of occurrence of 28, 14 y 7{\%} respectively.
This different probability of occurrence for the same structure with only different symmetries is due to rotational entropy, that also is responsible for those slight energy differences shown in Table~\ref{energia} and, in turn, it is the reason for the differences in the probability. At temperatures above 300 K, the probability of occurrence of the putative global minimum at cold temperatures
and depicted in solid-black line decay exponentially up to 1900 K. The dominant transformation solid-solid point ({\large{T$_{\textrm{ss$_1$-g}}$}})
is located at 739 K with 16.6{\%} of probability. 
%%%
%%%
%%%
At this point, there is a co-existence of chiral structures and achiral structures,  shown in Figure~\ref{geometry_gibbs}(a,g), above this point the achiral  structure (Figure~\ref{geometry_gibbs}(g)) become dominant. 
Its probability of occurrence is depicted in the solid-green line in Figure~\ref{popu}a and start to grow up at almost at room temperature.
The second transformation solid-solid point located at 1017 K and 10{\%} of probability also coexist the chiral putative global minimum with symmetry C$_1$ and achiral structure (Figure~\ref{geometry_gibbs}(h)) located at 2.40  kcal/mol Gibbs free energy at 298.15 K above the putative global minimum. Figure~\ref{popu}b shows the computed probability of occurrence considering the  percent enantiomeric excess is one hundred percentage, which implies of a pure Boltzmann ensemble of only one type of enantiomer. With the aim to compute the Boltzmann VCD/IR weighted spectra as a function of temperature, we take the relative population shown in Figure~\ref{popu}a. The probability of the dominant achiral putative global minimum with symmetry C$_1$ is depicted in the solid-green line in Figure~\ref{popu}a.  Analysis of the probability of occurrence leads to an interesting observation: The individual putative minimum global strongly dominates the VCD/IR at a
temperature ranging from 20 to 1240 K. The achiral structures have a zero contribution to VCD in hot temperatures. The probability of occurrence is dependent on the functional and basis set employed as a result of those energies computed at differents methods provides different energies~\cite{Puente} Figure~\ref{tspp} shows the relative population computed at TSPP~\cite{Tao}-GD3/def2-TZVP level of theory. At cold temperatures, strongly dominate the chiral structure with symmetry C$_1$ and depicted in   Figure~\ref{geometry_gibbs}(a,b). At hot temperatures, the dominant structure is a chiral helix-type structure depicted in Figure~\ref{geometry_gibbs}(k,l) located at 4.70 kcal/mol Gibbs free energy above the putative global minimum.  Also, at SP CCSD(T) level,  it is located at high energy above the global minimum. The relative population employing this functional does not follow the trend of energetic ordering at CCSD(T) level of theory. The above discussion shows that the probability of occurrence is sensitive to the functional. 
%%%
\subsection{Enantiomerization Energy Barrier at Finite Temperature.}
\begin{figure}[ht!]
\begin{center}  
  \includegraphics[scale=0.75]{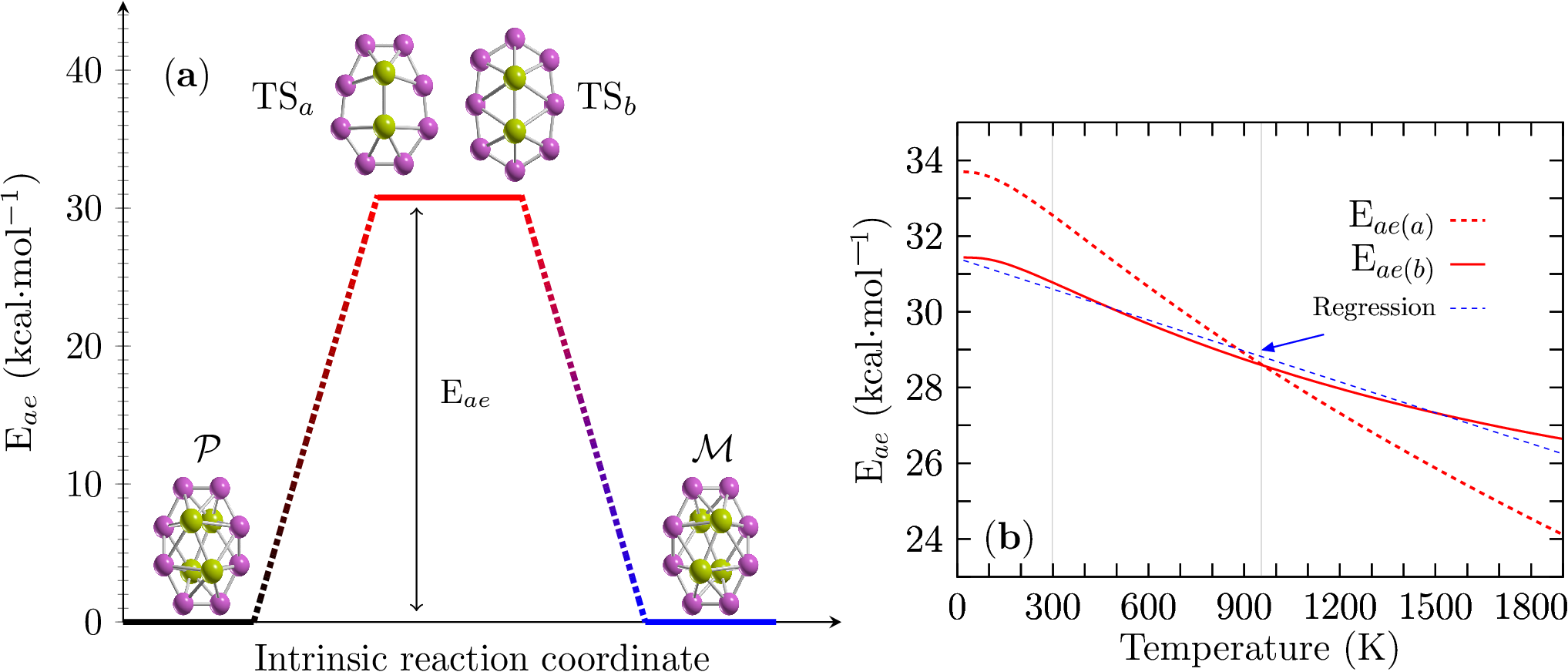}
  \caption{  (Color online) On panel (a), There are two transition states close in energy, TS$_a$ and Ts$_b$ depicted in
    panel (a) and two possible intrinsic reaction coordiantes (see movies in Supplementary Material):
    Route(a): Computed activation energy (E$_{ae(a)}$ ) of
    32.5  kcal·mol$^{-1}$.  Route(b): Computed activation energy (E$_{ae(b)}$)  of 30.77 kcal·mol$^{-1}$.
    between P and M chiral Be$_4$B$_8$ clusters at room temperature.
    The energy barrier for interconversion is equal to each other at
    temperature T, (Temperature barrier-barrier T$_{bb}$ point), which implies the velocity of reaction of
    both mechanism are equal. For the case of chiral Be$_4$B$_8$  T$_{bb}$ is located at 954 K.
    In panel (b) The E$_{ae(a)}$ is depicted in red dashed  line,  The E$_{ae(b)}$ is depicted red solid line for
    temperatures ranging from 20 to 1900 K.At 954 K.  The blue-dashed line on panel (b) represents a regression linear in the computed data of the energy barrier E$_{ae(b)}$ depicted in the red solid line in panel (b) its correlation coefficient is -0.9925. Harmonic frequency calculations were employed to make sure that the optimized transition states possess exactly one imaginary frequency.}
\label{barrier}
\end{center}
\end{figure}
The process in which pair of chiral molecules or enantiomers 
undergo to the conversion of one enantiomer into the other is referred to as enantiomerization.
Enantiomers each have the same free energy and equal probability of occurrence, as shown in Figure~\ref{popu}
The extent of interconversion of enantiomers depends on the energy barriers to enantiomerization. Moreover,
this energy barrier determines if an enantiomer can be resolved at temperature T and defines its configurational stability.
There are cases where the enantiomerization is undesirable; for example, many drugs are related to chirality, and frequently
only one of the enantiomers shows the desired effect while the other shows undesirable effects, moreover chiral molecules with
high charge-carrier mobility and fluorescence quantum yield needs high energy barriers of enantiomerization.~\cite{Ravat}

Figure~\ref{barrier}a shows the computed  enantiomerization energy barrier (energy activation ($E_{ae}$) or Gibbs free activation energy ($\Delta$G$^{\ddagger}$))  of the pair of enantiomers $\mathcal{P}$ and $\mathcal{M}$ of Be$_4$B$_8$ cluster that has only a single-step for two mechanisms of reaction whose energy barriers are energetically similar.  The transition states (TS$_a$, TS$_b$) depicted in  Figure~\ref{barrier}a,  are achiral sandwich-type structures in which the borons form a planar ring with each of the Be-Be dimers capping the top and bottom, and thery are aligned parallel to the major axis of the boron ellipse. The main difference between both of them is a shift of ring position concerning the Be-Be dimers.  The energy barriers of TS$_a$ and TS$_b$ are 32.50 and 30.77 kcal/mol, respectively, and indicates that Be$_4$B$_8$ enantiomers are stable at room temperature. Those energy barriers height are similar to that of computed energy barrier height in Au$_{38}$(SR)$_{24}$,~\cite{Malola} clusters that lie in the range of 29.9 to 34.5 kcal/mol.  The energy of enantiomerization, E$_{ea(a)}$ and E$_{ea(b)}$ corresponding to the TS$_a$ and TS$_b$ for temperatures ranging from 20 to 1900 K are displayed in  Figure~\ref{barrier}b respectively. The E$_{ea(a)}$  is depicted in a  red dashed line, whereas the  E$_{ea(b)}$ is depicted in a red solid line. Analysis of the results leads to an interesting observation: In Figure ~\ref{barrier}b, one can see that there is a  barrier-barrier temperature point (T$_{bb}$) located at 954 K  where the energy barriers of both mechanisms are equal to each other. At T$_{bb}$, the probability that reaction takes one path or another is 50/50{\%}  which implies that the velocities of reaction for both reaction mechanism are equal to each other.  Below the temperature of 954 K, the reaction path b (TS$_b$) is more favorable than reaction path a (TS$_a$), and vice-versa for temperatures above 954 K.  
In Figure~\ref{barrier}b. is shows that the energy barrier E$_{ea(b}$ decreases linearly in the temperature
range from 200 to 740 K. Below 200 K and in temperatures ranging from 740 to 1900 K; the energy barrier behavior is non-linear.
To make it clearer, a line depicted in the blue dashed line in Figure~\ref{barrierR} appendix~\ref{tomate} overlapping to energy barrier in the temperature range 200 to 740 K. Equation~\ref{tempe} was found by the linear regression, with correlation coefficients -0.9925, of the energy barrier depicted in a red solid line in Figure~\ref{barrier}b.
\begin{equation}
\displaystyle 
\Delta G^\ddagger=31.41-0.00271188\cdot T
\label{tempe}
\end{equation}
In Equation~\ref{tempe}, T is the temperature, and it describes approximately the energy barrier in all ranges of temperature. It is depicted in the blue dashed line of Figure~\ref{barrier}b. Evaluating Equation~\ref{tempe} with T=298.15 K gives 30.59 kcal/mol,  just very close to the computed value of 30.77 kcal/mol.  The first term of the Equation~\ref{tempe} is enthalpy, and the second one is the entropic term. The computed values of $\Delta$G, $\Delta$H, $\Delta$S, and the percentage of contribution of $\Delta$S to energy barrier are summarized in Table~\ref{entro} for some temperature values.
Analysis of results shown in Table~\ref{entro} leads that the enthalpy term is too large compared with the entopic term shown in row 3 and row 4 of Table~\ref{entro}, respectively, and evaluated in ranging temperatures given in row 1 of Table~\ref{entro}. In row five of Table~\ref{entro} is given the percentage in which the energy barrier decreases as a function of temperature and due to entropic term considering as reference one hundred percent when the energy barrier is computed at T=0. Notably, the composition of the energy barrier isenthalpic and is too high in all ranges of temperature. We concluded that the interconversion between enantiomers is thermodynamically unfavorable in all ranges of temperature based on our computations. At hot temperatures, the energy barrier still is too high, and the most significant entropic contribution is not more than 15.54{\%}. Similar results are obtained for the TS$_b$. 
\begin{table}[!ht]\centering
  \caption{ Approximate, energy barrier, enthalpy, entropy terms, and 
    the percentage in which the energy barrier decreases, considering the reference of one hundred percent when the energy barrier is computed at T=0. The lowering of the energy barrier is due to entropic terms.($\Delta$S$\cdot$T)     
The highest value is  26.52{\%} at hot temperatures, this leads to the observation that the composition of the barrier is at least is {84.4\%} enthalpic in all ranges of temperature.}
  \label{entro}
   \begin{tabular}{@{\extracolsep{0.2pt}} cc c cc}
 \\[-1.8ex]\hline 
\multicolumn{1}{c} {{{ \normalsize{Temperature (K) }}}}  & \multicolumn{1}{l}{$\Delta$G} & \multicolumn{1}{c}{$\Delta$H} & \multicolumn{1}{c}{$\Delta$S$\cdot$T } & \multicolumn{1}{c}{ {\%}(Decrease) }\\
\hline \\[-2.2ex]
0                         &31.41  & 31.41 &  0.0      &  0.0 \\
300                       &30.60  & 31.41 & -0.81     &  2.58 \\
600                       &29.79  & 31.41 & -1.62     &  5.17\\
900                       &28.97  & 31.41 & -2.44     &  7.76\\
1200                      &28.16  & 31.41 & -3.25     &  10.35\\
1500                      &27.35  & 31.41 & -4.06     &  12.94\\
1800                      &26.53  & 31.41 & -4.88     &  15.53\\
\hline \\[-1.8ex]
\end{tabular}
\end{table}
\subsection{Energy Barrier beteween Chiral and Achiral Structures at Finite Temperature.}
\begin{figure}[ht!]
\begin{center}  
\includegraphics[scale=0.75]{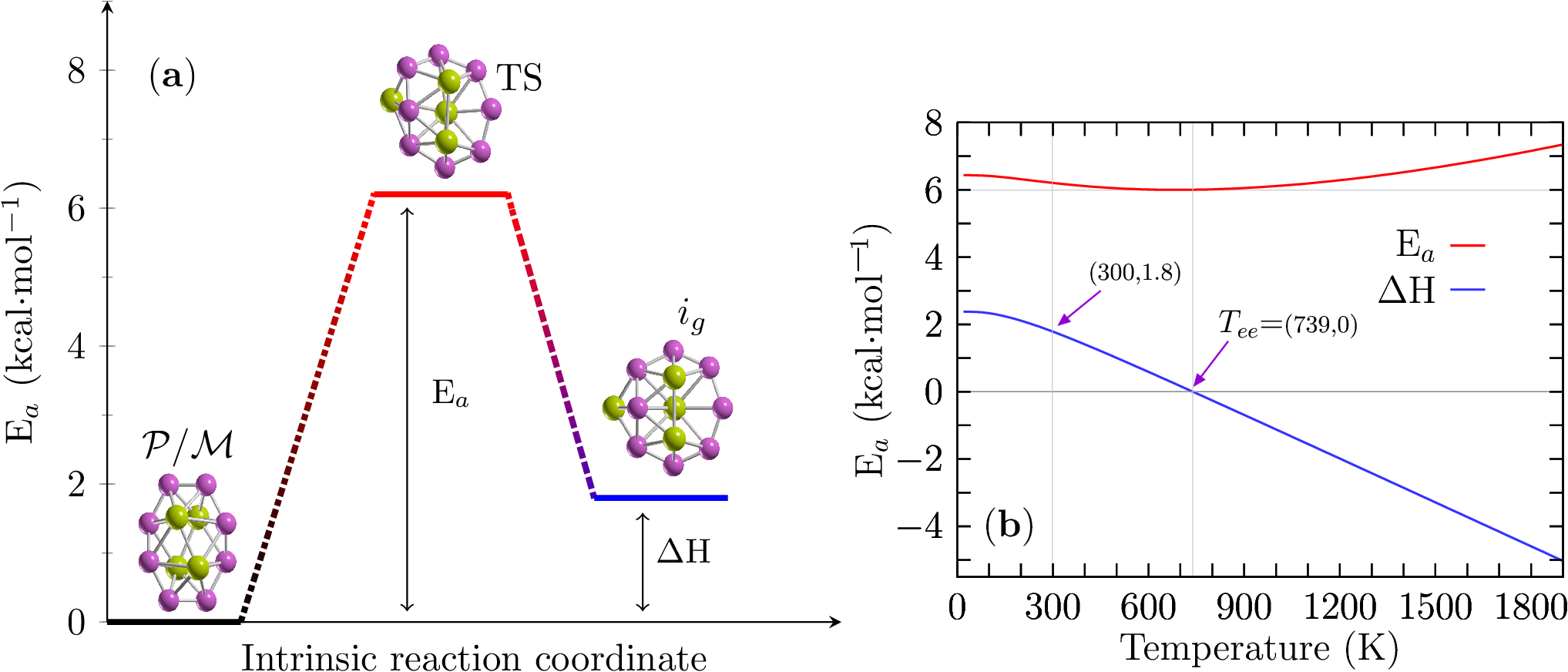}
\caption{(Color online) On panel (a), The height of energy barrier for interconversion (E$_a$) is 6.20 kcal$\cdot$mol$^{-1}$ between $\mathcal{P}$  or $\mathcal{M}$  chiral and achiral structures (see Figure~\ref{geometry_gibbs}7) of  Be$_4$B$_8$ clusters at room temperature. On panel (b), the  E$_a$ is depicted in the red solid line, and the enthalpy of formation ($\Delta$H) depicted in blue solid line for temperatures ranging from 20 to 1900 K. The $\Delta$H is zero at T=739 K and 1.8 kcal/mol at T=300 K. Notice, E$_a$ has a minimum located at 739 K.}
\label{barrier2}
\end{center}
\end{figure}
Figure~\ref{barrier2}a displays the height of energy barrier interconversion at room temperature between the chiral
 $\mathcal{P}$/$\mathcal{M}$ structures shown in Figure~\ref{geometry_gibbs}(1,2), and achiral structure depicted  in Figure~\ref{geometry_gibbs}7.  Remarkable, these structures coexist at the dominant solid-solid transformation point located at 739 K, and according to the probability of occurrence,  at hot temperatures, the achiral structure is the putative global minimum. Furthermore,
the endergonic to an exergonic temperature point, T$ee$, is defined here as the temperature where
the reaction type change from an endergonic to an exergonic and in this Be$_4$B$_8$ clusters it coincides with the solid-solid
transformation point.  When these two points coincide,  at least two structures coexist, and there
is a change of type reaction from endergonic to an exergonic or vice-versa. 
For the interconversion between those structures, the height of the energy barrier at room temperature is 6.20 kcal/mol,
and the enthalpy of formation (AH) is 1.8 kcal/mol. The TS is depicted in  Figure~\ref{barrier2}a, it is also a sandwich type
structure formed by a distorted circular ring in which the Be-Be dimers are capping each faces of the ring, it has similar
structure to isomer
$i_7$ depicted in Figure~\ref{geometry_gibbs}(g). Figure~\ref{barrier2}b shows the height of the energy barrier for chiral
and achiral structures depicted in a red solid line, the enthalpy of formation ($\Delta$H)) for the same structures is
depicted in a solid blue line for temperatures ranging from 20 to 1900 K. An analysis of $\Delta$H in Figure~\ref{barrier2}b,
shows that the reaction process is endothermic for temperatures ranging from 20 to 739 K because the $\Delta$H is positive.
At the temperature of 739 K, the $\Delta$H is zero,  which implies that the chiral structure with C$_1$ symmetry and
achiral ($i_7$) structure with C$_1$ symmetry coexist. The above discussion is in good agreement with the computed
point T$_{ss-1}$ located at 739 K displayed in Figure~\ref{popu}a,  and according to the probability occurrence, at
this point, the chiral and chiral structures coexist.  Furthermore, in this temperature point, the height of the
energy barrier, depicted in the red-solid line Figure~\ref{barrier2}b, has a minimum value of 6.0 kcal/mol.  At
temperatures above 739 K, the reaction process is exothermic due to the $\Delta$H is negative, and the height of
the energy barrier starts to increase slowly. Analysis of results in more detail leads to several observations.
The reaction process is endothermic up to 739 K,  which implies the absorption of energy, and the chiral structures
strongly dominate as the putative global minimum.  At temperatures of 739 K, the chiral and not chiral structures coexist.
At temperatures above 739 K,  the reaction process is exothermic, which implies a heat of reaction, and the non-chiral
structures weakly dominate as the putative global minimum.  Based on the  $\Delta$H  behavior in all ranges of temperature,
we point out that the reaction is an entopic-driven process due to that type of reaction change from endothermic to exothermic
as the temperature increases.
\begin{figure}[ht!]
\begin{center}  
  \includegraphics[scale=0.80]{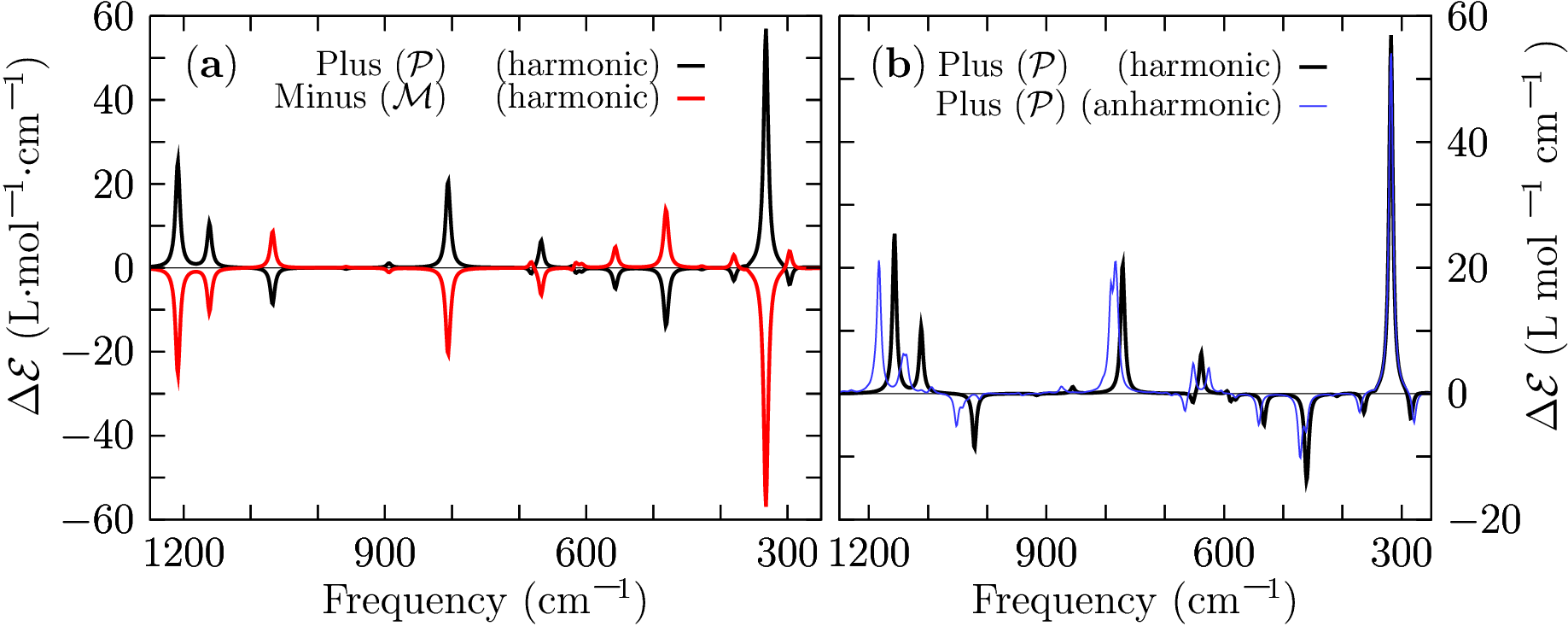} 
\caption{Panel (a) Shows a comparison of the VCD spectrum for frequencies ranging from 1250-250 cm$^{-1}$, of the lowest energy $\mathcal{P}$ and $\mathcal{M}$ enantiomers and in Panel (b) it shows a comparison of the VCD-harmonic spectrum  and VCD-anharmonic spectrum to estimate the importance of anharmonicities of the Be$_4$B$_8$ chiral cluster. The anharmonic vibrational spectrum is depicted in the solid blue line, together with the harmonic vibrational spectrum depicted in solid black line for the for the lowest energy enantiomers $\mathcal{P}$. A scale factor of 0.96 is applied to shift the harmonic spectrum to overlay on anharmonic spectrum. The spectra are results of a convolution  of a Lorentzian  shape profile  with FWHM of 20 cm$^{-1}$ with the computed discrete frequency intensities. The x-axis is given the frequency in cm$^{-1}$ and in y-axis is given in units of molar absorptivity ($\Delta\mathrm{\mathcal{E}}$)}
\label{armo}
\end{center}
\end{figure}
\subsection{VCD and IR spectra}
Figure~\ref{armo}a. shows a comparison of the VCD harmonic spectra corresponding to $\mathcal{P}$ and $\mathcal{M}$ lowest energy structures depicted in solid black- and red lines, respectively. They show a mirror image relationship, thereby ensuring that the two structures are non-superposable. The computed VCD spectrum ($\mathcal{P}$ structure) is characterized by five main peaks located at frequencies of 330 cm$^{-1}$, 481 cm$^{-1}$, 802 cm$^{-1}$,  1062 cm$^{-1}$, and 1208 cm$^{-1}$ respectively. The largest peak positive intensity is located at 330 cm$^{-1}$, and it corresponds to the stretching of the two Be-Be dimers that capped the distorted boron ring. Next, a transition located at 481 cm$^{-1}$ is the largest negative and is attributed to bending of the boron distorted ring, a kind of breathing motion. The peaks located in the region 1062 cm$^{-1}$ to 1208 cm$^{-1}$ corresponds to ring boron stretching. The harmonic approximation works if the potential energy is parabolic and it fails~\cite{Glaesemann} as the temperature increases due to anharmonic effects~\cite{Glaesemann} Under harmonic approximation, strongly anharmonic systems are not well described~\cite{Hellman,Grimvall} For high temperatures above 0.7...0.8 melting temperature, explicit anharmonic contributions become relevant,~\cite{Neugebauer} moreover, we have to consider if the cluster whether or not it is highly strongly anharmonic.
\begin{figure}[ht!]
\begin{center}  
\includegraphics[scale=0.80]{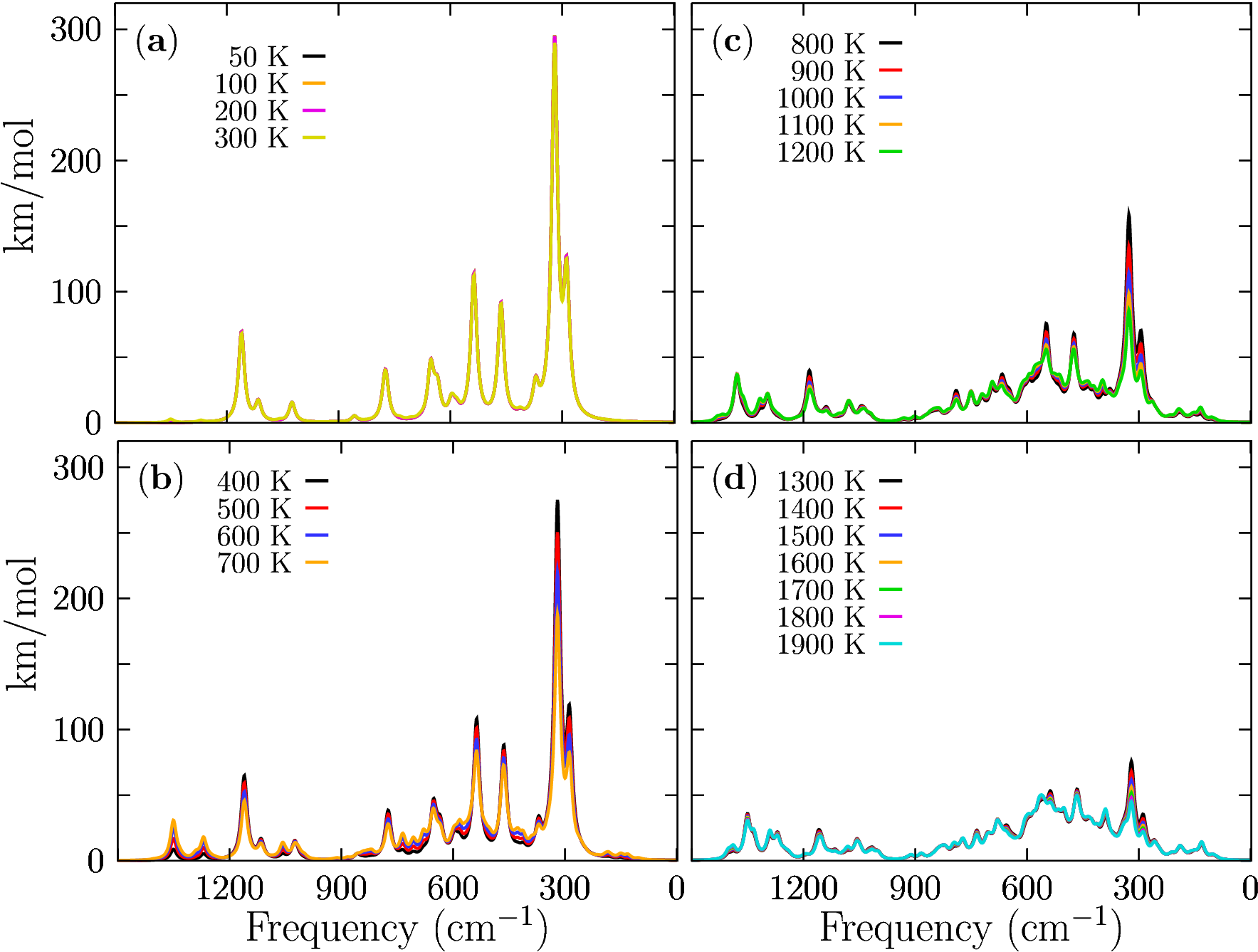}
\caption{The dependent temperature IR Boltzmann-spectra-weighted  of the $\mathcal{P}$ Be$_4$B$_{8}$ ensemble computed at the PBE0-D3/def2-TZVP level of theory computed in frequency range of 1500 to 1 cm$^{-1}$. Three similar chiral conformers with C$_1$, C$_2$ and D$_2$ symmetries which correspond to 96.3{\%} of the Boltzmann distribution, strongly dominate the IR Boltzmann-spectra-weighted in temperature form 0 to 1200 K. The IR Boltzmann-spectra-weighted for temperaturas ranging: in panels (a) 50-300K, (b) 400-700, (c) 800-1200, and (d) 1300-1900 K. At temperatures below 300 K, the amplitud of spectra are constant in good agreement of the realtive population. At temperature above 300 K the magnitude of the spectrum decreases exponentially until to 1200 K. The spectra was computed employing a Lorentzian with half-widths at half-maximum of 20 cm$^{-1}$. The computed frequencies were multiplied with a scaling factor of 0.96. The x-axis is given the frequency  and y-axis is IR intensities in km/mol.}
\label{ir2}
\end{center}
\end{figure}
To estimate the importance of anharmonicities of the Be$_4$B$_8$ chiral cluster, we show, In panel (b) of Figure~\ref{armo}b the anharmonic VCD spectra that is depicted in solid blue line, and for ease comparison overlaying  with the harmonic vibrational spectrum depicted in solid black line, both of them computed for $\mathcal{P}$ lowest energy structure, and employing Gaussian code.~\cite{gauss} A shifting factor of 0.96 is applied to shift the harmonic spectrum to overlaying the anaharmonic spectrum. We found that the frequency shift is 14 cm$^{-1}$ towards to high energy. A comparison of two spectra dispalyed in Figure~\ref{armo}b shows that The computed harmonic and anharmonic spectra are in very good
agreement. In fact, most of the peaks are correctly computed employing the harmonic approximation. In the low range of energy the harmonic peaks and anharmonic peaks agree well, however, there are slightly discrepancies in the region 1100 cm$^{-1}$ to 1200 cm$^{-1}$, nevertheless, in this study the computations of the thermodynamic properties and VCD spectra under the harmonic approximation yield
reliable results enough to describe the non-strongly anharmonic Be$_4$B$_8$ chiral cluster. Additionally, the Figure~\ref{irspectra}, appendix~\ref{irspectra2} show the IR spectra computed under harmonic and anharmonic approximations.
The IR-harmonic spectrum is depicted in solid black line, whereas, the IR-anharmonic spectrum is depicted in solid red line. A shifting factor of 0.96 is applied to match the IR-harmonic spectrum over the IR-anharmonic spectrum. Comparing both spectra, it can see that the spectra match over a large range of frequencies.  From the above mentioned we infer that the IR spectra under harmonic approximation yield valid results.
Regarding dependent-temperature VCD spectra, the  Boltzmann weighted overlapping to yield a total VCD spectrum at all ranging temperatures is zero because the Boltzmann ensemble is composed of achiral structures and an equal mixture of both $\mathcal{P}$ and $\mathcal{M}$ enantiomers, which implies that the Boltzmann ensemble is racemic. Any chiroptical response in the Be$_4$B$_8$ cluster is going to be null. The exhaustive exploration of potential and free energy surface revealed that there are twenty two isomers within an energy range of 9.2 kcal/mol, six of which are chiral structures, with symmetries C$_1$, C$_2$, D$_2$ respectively, and were within 1 kcal/mol. Moreover, these structures compose 98{\%} of relative population at room temperature. With the aim to compute  Boltzmann weighting IR spectra, those structures which only differs on symmetries has to be taken into account The IR spectra, in comparison to VCD spectra, is not null.
Figure~\ref{ir2} shows the IR spectra for temperatures ranging from 50 to 1900 K.
The  IR spectrum is composed of five peaks. The largest peak intensity is located 330 cm$^{-1}$  frequency axis it corresponds to the alternating stretching of the two Be-Be dimers
that capped the distorted boron ring, and it is a mode that contributes to interconversion between P and M structures, The other four modes are related to compression/expansion of the boron ring.

Figure~\ref{ir2}a display the IR spectra for temperatures ranging from 50 to 300 K, in this
range, the IR spectra is strongly dominated by the spectrum of the lowest energy pair of
enantiomers with C$_1$ symmetry, and further the IR intensities remain constants in all
range of temperatures. The above mentioned, is in agreement with the relative
population depicted in Figure~\ref{popu}a where the probability of occurrence of the
pair of enantiomers with symmetry C$_1$ strongly dominate. We have to consider that 
the contribution to IR spectra of the four enantiomers with C$_2$ and D$_2$ symmetries
for temperature ranging from  20 to 300 K is equal to the IR spectrum with symmetry C$_1$, and
there is not presence of other structure. Thus at room temperature
all molecular properties, except for the chiral properties, are attributable to the lowest energy pair of enantiomers, depicted in Figure~\ref{geometry_gibbs}(a,b). Panel (b) of Figure~\ref{ir2} shows the IR spectra
for temperature ranging from 400 to 700 K. The IR intensities start with exponential decay, in agreement
with the probability of occurrence of the lowest pair of enantiomers of Figure~\ref{popu}a.  There is a little contribution from other isomers, but not enough to change/alter the IR spectrum. Thus the shape of IR spectrum remains equal to  IR spectrum at cold temperatures. The IR spectra for temparatures ranging from 800 to 1200 K are shown in Figure~\ref{ir2}c, the largest contribution of a particular isomers is less than 17{\%}
thus the largest peak of IR spectra trend to be neglected. Panel (d) in Figure~\ref{ir2}
displays that the IR spectra is almost null, thus at hot temperatures the IR spectra
is neglected becuasen of almost all contributions of the isomers to IR spectum are around 10{\%}.
\subsection{Molecular Dynamics}
We perfomed a Born-Oppenheimer molecular dynamics employing the  deMon2K program~\cite{demon2k}  (deMon2k v. 6.01, Cinvestav, Mexico City 2011) at different temperatures (1200, 1500 and 1800 K), with the aim to gain insights into the dynamical behavior of Be$_4$B$_{8}$ cluster. (See movies in supplementary material)  We employed similar parameters that in a prevoius work.
The simulation time was of 25 ps with a step size of 1 fs. For Be$_4$B$_{8}$ cluster we found a dissociation phenomena when the temperature is higher than 1500 K, at 1800 K the dissociation process is stronger, while at 1200 K there is no dissociation. When at temperature T,  a cluster dissociate the melting point temperature is lower that temperature of   dissociation.~\cite{Li-Min,Buelna}.

\section{Conclusions}
For the first time, to our knowledge and from our results,
we found that the chiral cluster Be$_4$B$_{8}$ is the lowest energy structure in ranging temperatures from 0 to 739 K. 
We initially sampling the potential energy surface
with 2400 candidates of neutral Be$_4$B$_{8}$ structures, and with an efficient cascade
type algorithm coupled to DFT; We were able to locate twenty-two final isomers within an
energy range of 9.2 kcal/mol, six of which are chiral low energy structures with symmetries
C$_1$, C$_2$, and D$_2$ respectively, as far as we know not yet reported as chiral structures. 
They are constituted by a sandwich structure type in
which the boron atoms from a distorted ellipsoid ring with each Be-Be dimers are capping each
side of the sandwich. Our findings show that the chirality that exhibits the Be$_4$B$_{8}$ cluster
emerges from the Be-Be dimers' mirror position. Additionally, based on the AdNDP analysis and the computed transition state
and IRC (see movie Supplementary Information)  between $\mathcal{P}$ and $\mathcal{M}$ enantiomers,
we can deduce that the Be-Be and Be-B
interaction favors the Be$_4$B$_8$ to be chiral and energetic minimum structure. According to our
calculations, the enantiomerization energies barriers of the transformation $\mathcal{P}$ to
$\mathcal{M}$ and $\mathcal{M}$ to $\mathcal{P}$ is equal to each other, moreover  the activation
energy between $\mathcal{P}$ to an achiral structure and $\mathcal{M}$ to an achiral structure are
equal, thus the Boltzmann ensemble is composed of an equal
mixture of $\mathcal{M}$ and $\mathcal{P}$ enantiomers.
The enantiomerization energy between
the chiral putative global minima in the ranging of temperatures from 20 to 1900 K is mainly composed
of enthalpic than entropic term. The entropy-temperature term reduces the interconversion energy
barrier like maximum in 28{\%}; thus, the barrier is not influenced significantly by the entropic
term. On the other hand, our results show that the energy barrier is 6.20 kcal/mol at room temperature
between the lowest energy enantiomer ($\mathcal{P/M}$ and the first achiral structure
located at 1.79 kcal/mol at
298.15 K. The barrier does no increase significantly neither at cold temperatures nor at hot
temperatures, the barrier high at 20 K is 6.43 kcal/mol. In contrast, it is 7.32 kcal/mol at 1900 K.
Remarkably, the interconversion between enantiomers and achiral structure is more favored than the
interconversion between a pair of enantiomers in all ranges of temperature.
Furthermore, our results show that in the temperature range from 20 to 738 K, the interconversion between
an enantiomer ($\mathcal{P/M}$ and the achiral structure is an endothermic type reaction. At the
transformation solid-solid point located at 739 K, the enthalpy of reaction is zero,
and the energy barrier is a minimum. In this point coexist the  $\mathcal{P}$
and $\mathcal{M}$ structures with an achiral structure.
At 739 K, above the interconversion between enantiomers-achiral
structures are exothermic, which implies the heat of reaction. Notice that the change from endo to exothermic reaction is driven by entropic-temperature term. In this study and for the first time, we defined two novel points on a scale of temperature. a) The first one is the T${_{ee}}$ where the change in the reaction type occurs, and b) the second is the  T${_{bb}}$ where the activation energy of two different reaction mechanisms are equal to each other, so the reaction rate equals.
The energetic ordering of 
the isomers when we employing the Gibbs free energy computed at 298.15 Kagree with the energy ordering
employing the single point CCSD(T) energies computed with DLPNO-CCSD(T), and the energy ordering change
just with the interchange of the two achiral structures located at 3.61 kcal/mol and 3.38 kcal/mol above
the putative global minimum. Furthermore, Our results on $\mathcal{T}_1$ diagnostic confirm that energies
computed at the DFT level of theory do not contain a large multireference character, so the
Be$_4$B$_8$ cluster is well characterized. Regarding relative population, our findings show that the
putative global minimum that clearly and strongly dominates at cold temperatures is the pair of enantiomers.
At 739 K, above, an achiral structure dominates as a putative global minimum with less than 20{\%}.
We also show that the relative population or Boltzmann distribution could depend on the functional and
the basis set employed. The energy separation among isomers affects the relative population strongly.
The Boltzmann distribution is composed of six isomers at 298.15 K with only different symmetries,
and its contribution is approximately  98{\%} thus, at room temperature, the molecular
properties are strongly dominated by the pair of enantiomers putative global minimum. We analyzed
the effects of the point group symmetry on Gibbs free energy as a function of temperature.
The small Gibbs free relative energy  differences of 0.41 and 0.81 kcal/mol between different
symmetries at 298.15 K are due to rotational entropy that in fact, it is a function of number of
symmetry, and it become larger at the temperature increase.
This strongly suggests that we need to explore the free energy surface
taking into account clusters with low symmetries. The low
symmetries trends to dominate as putative global minimum as the temperature increases. 
For total VCD/IR spectra as a function of
temperature, we compute the Boltzmann weighted superposition of each isomer's VCD/IR spectrum that
yields a total VCD/IR spectrum. Boltzmann VCD weighted spectra for this particular chiral cluster is
null in all range of temperature because in the ensemble P and M enantiomers
are present in equal mixture.  
However, a clear
temperature dependence of the Boltzmann IR weighted spectra is driven just by the probability of
the putative low-energy isomers in the temperature ranging from 20 to 739 K. At temperature above
739 K, the IR spectra decay strongly, whereas, at a temperature above 1200 K, the IR spectra are
almost null. In summary, IR spectra at room temperature are dominated by pair of enantiomers putative
global minima, whereas at hot temperatures, the IR spectra are almost null. Our molecular dynamics
corroborate that the melting point is located in the temperature ranging from 1400 to 1500 K,
because and in accorde with molecular dynamics, the dissociation phenomenon occurs at 1500 K, and the
melting point is located below this point. As future work, the inclusion of anharmonic effects
should be taken into account also the relative population at CCSDT(T) level of theory is going to
be calculated. The  melting point also should be computed. At finite temperature the entropy is
maximized, taking into account this, a future project is a implementation  of an algorithm to search
on the free energy surface employing the entropy as an objective function instead of energy. Finally,
we point out that in any reaction, the activation energy at finite temperature must be computed
considering the reactants and products with non-symmetry (C$_1$), by reason of the small energy
differences dependent on symmetry and temperature (i.e 0.41 kcal/mol for C$_2$ at 298.15 K)
increase the energy barrier between the reactants and products, hence the velocity of the reaction
could be miscalculated. The Boltzmann Optics Full Adder (BOFA) \emph{Python} code supporting
the findings of this study is available from the
corresponding author upon reasonable request.
\section{Acknowledgments}
C. E. B.-G. thanks Conacyt for the Ph.D. scholarship (860052). E. R.-CH. thanks Conacyt for the Ph.D. scholarship (1075701). We are grateful to Dra. Carmen Heras, and L.C.C. Daniel Mendoza for granting us access to their clusters and computtional supporting. Computational resources for this work were provided by \emph{ACARUS} through the High-Performance Computing Area of the University of Sonora. Sonora, M\'exico. We are also greatful to the computational chemistry laboratory for providing computational resources, \emph{ELBAKYAN} and \emph{PAKAL} supercomputers. Powered{@}NLHPC: This research was partially supported by the supercomputing infrastructure of the NLHPC (ECM-02) in Chile.
\section{Conflicts of Interest} The authors declare no conflict of interest.
\section{Funding} This research received no external funding.
\section{Supplementary Materials}
The following are available online at \href{https://youtu.be/x9eYp0oWmg0}{IRC route A} \href{https://youtu.be/W-0ObvALZIo}{IRC route B}.  All xyz atomic coordinates optimized of the Be$_4$B$_{8}$ cluster at the PBE0-D3/def2-TZVP/Freq.
\section{Abbreviations}
The following abbreviations are used in this manuscript:
\noindent 
\begin{tabular}{@{}ll}
  DFT & Density Functional Theory\\
  CCSD(T)  & Coupled Cluster Single-Double and perturbative Triple\\
  DLPNO-CCSD(T) & Domain-based Local Pair Natural Orbital Coupled-Cluster Theory\\
  ZPE & Zero-Point Energy\\
  VCD & Vibrational Circular Dichroism\\
  IR  & Vibrational Infrared Spectrum\\
  BOFA &  Boltzmann-Optics-Full-Ader code\\
  GALGOSON & Global Genetic Algorithm of University of Sonora\\ 
  SSh    & Sandwich-Structure-Hollow\\
  AdNDP  & Adaptive Natural Density Partitioning\\
\end{tabular}
\bibliography{bibliografia}
\newpage 
\appendix
\section{Probability of occurence computed at TPSS-GD3/def2TZVP level of theory.}\label{poput}
\begin{figure}[ht!]
 \begin{center}  
 \includegraphics[scale=0.80]{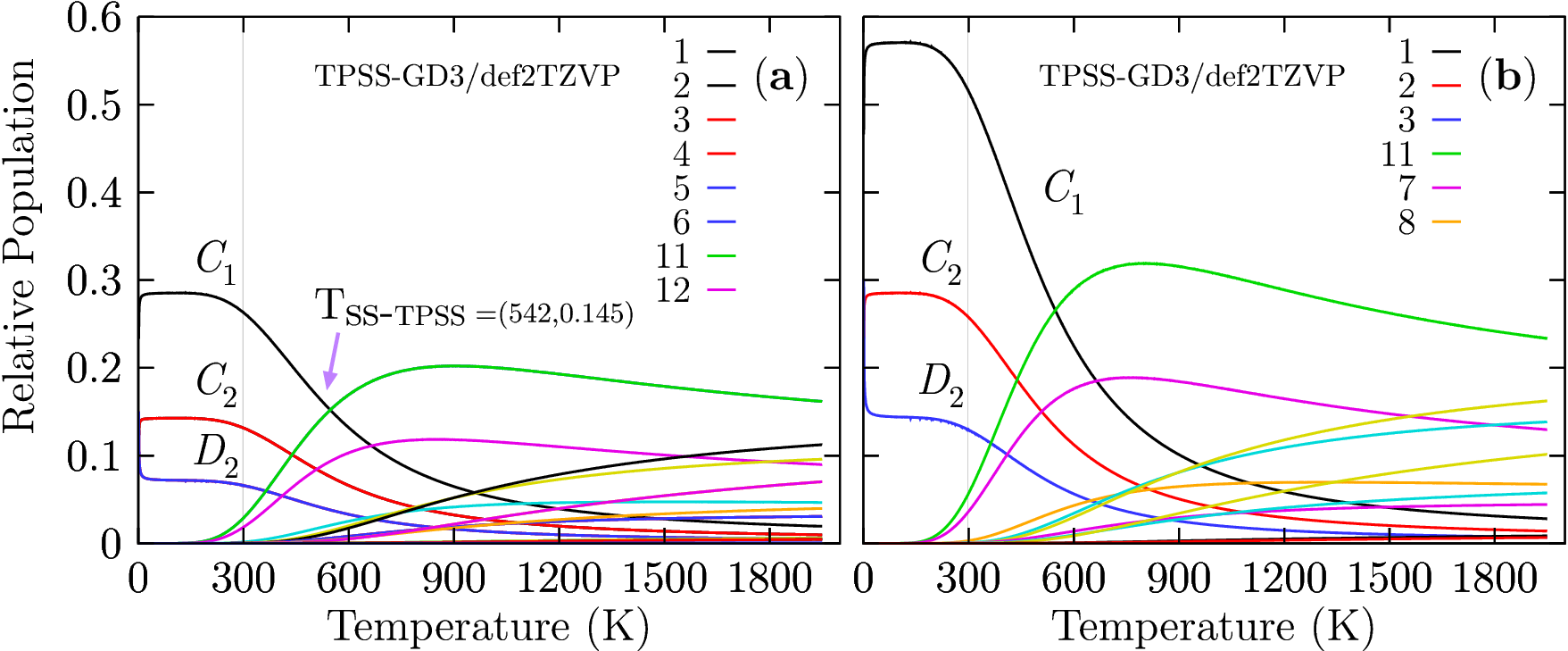}
 \caption{Probability ocurrence of each isomers computed employing TPSS~\cite{Tao} functional with the def2TZVP basis set, taking into account version three of Grimme's dispersion~\cite{Grimme} as it is implemented in Gaussian code.~\cite{gauss} The relative energies between two isomers vary considerably with the functional use. This will affect the temperature-dependent  Boltzmann factors computed for each isomer and, therefore, the relative population change, as shown in Figure. Employing TPSS functional.  The T$_{SS}$ point is located at 542 K on a temperature scale compared with the T$_{SS}$ point located at 739 K when we  employ the PBE0 functional.}
 \label{tspp}
 \end{center}
\end{figure}
For temperatures ranging from 20 to 542 K, the chiral structure, depicted in Figure~\ref{geometry_gibbs}(a), strongly dominates as a global minimum. At 542 K, a type-helix chiral structure coexists, depicted  Figure~\ref{geometry_gibbs}(k), with the chiral putative global minimum at cold temperatures. For temperatures ranging from 543 to 1600 K, the chiral structure's type-helix chiral structure dominates as the putative global minimum. A large difference between the TSS point computed with TPSS functional~\cite{Tao} and that computed with PBEO functional~\cite{Adamo}. Thus, it is important to choose the DFT functionals and basis that better describe the system to calculate the relative population and any Boltzmann weighted property.  
%%%%%%%
%%%%%%%
\newpage
\section{Be-Be and B-B bond length evolution for enantiomers}\label{BeBe}
\begin{figure}[ht!]
  \begin{center}  
    \includegraphics[scale=0.80]{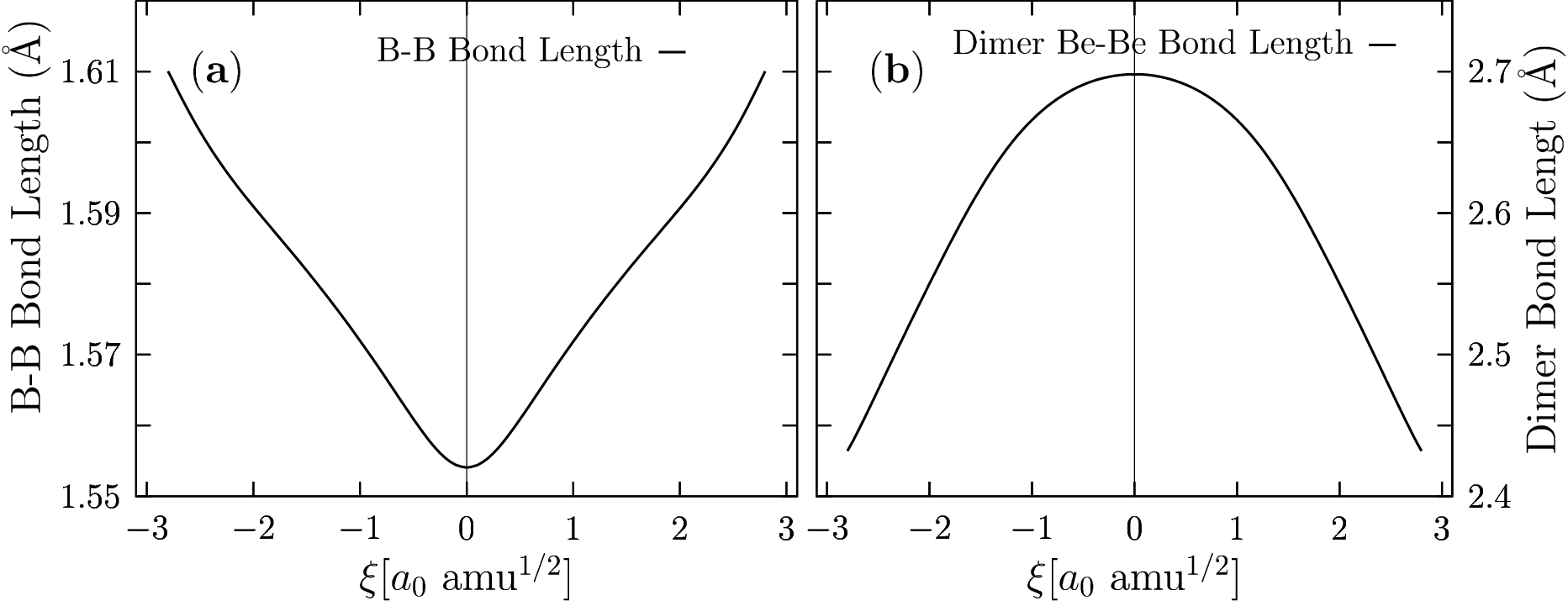}
    \caption{ Panel (a) shows the bond length evolution of the Be-Be dimer that is capping one side of the distorted ring boron along the IRC of the chiral Be$_4$B$_8$ cluster. Panel (b) shows the evolution distance between the two dimers that are capping the distorted ring boron along the IRC of the chiral Be$_4$B$_8$ cluster. In panel (a), the minimum Be-Be bond length is located at TS state with value of 1.9416~\AA,~and
      the maximum value is 1.9862~\AA that correspond to one of the putative global minimum.
      The largest rate of decreasing/increasing bond length of Be-Be dimer is happening when the reaction start/end, before or after
      of the point of maximum force.}
    \label{evo}
\end{center}    
\end{figure}
\newpage
\section{Energy  of enantiomerization}\label{tomate}
\begin{figure}[ht!]
  \begin{center}  
    \includegraphics[scale=0.80]{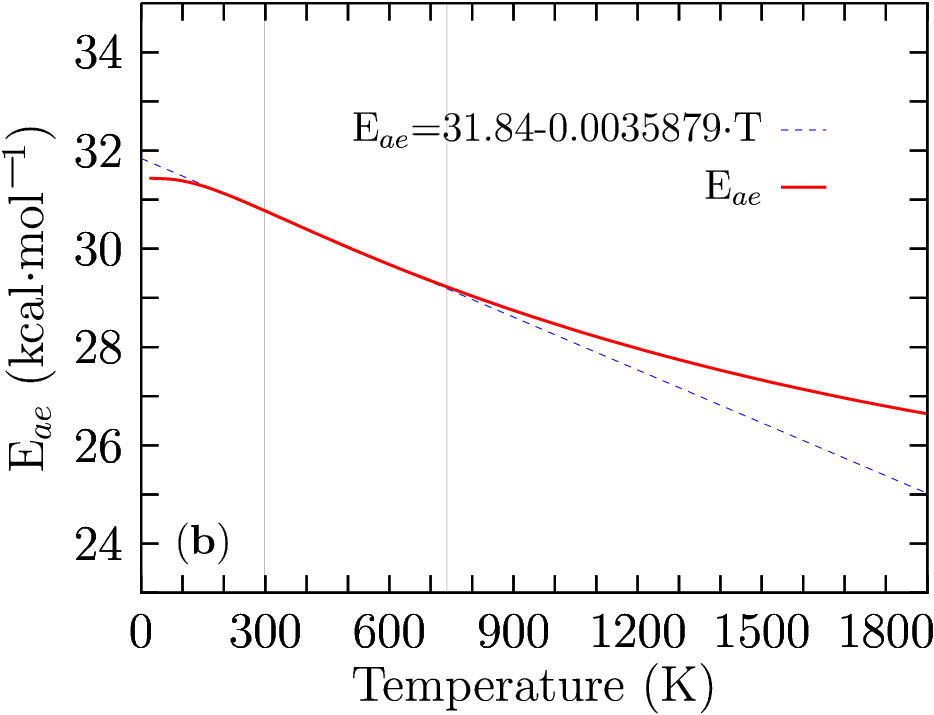}
    \caption{We show a perfect line depicted in the blue dashed line overlapping to energy barrier for enantiomers in the temperature range 200 to 740 K.}
    \label{barrierR}
\end{center}    
\end{figure}
%\newpage
\section{IR Harmonic vs Anharmonic spectra}\label{irspectra2}
\begin{figure}[ht!]
  \begin{center}  
    \includegraphics[scale=0.80]{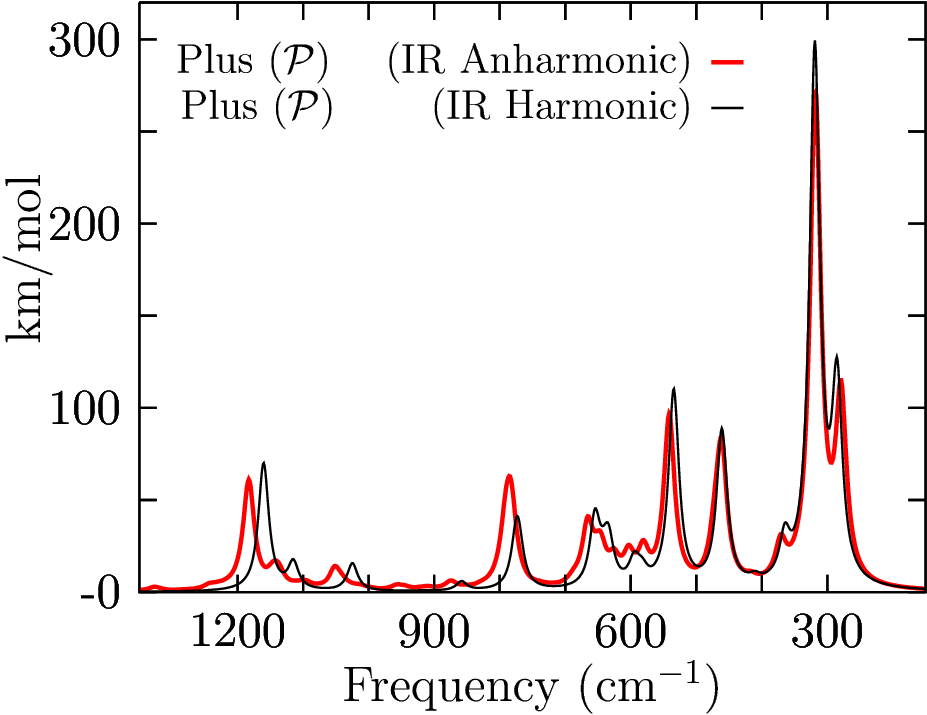}
    \caption{We show a comparioson between IR Harmonic vs IR Anharmonic spectra.
      IR-Harmonic spectrum was scaled by 0.96 to overlap the IR Anharmonic spectrum.
      The full width at half maximum (FWHM) employed is 20$^{cm-1}$}
    \label{irspectra}
\end{center}    
\end{figure}
\newpage
\section{XYZ atomic coordiantes}
\label{appendix:c}
\begin{verbatim}
12    
0.000000000         cluster_0002.out
Be  -0.905655000000  -0.409167000000  -1.143981000000
Be   0.906419000000  -0.409960000000   1.144242000000
Be  -0.905173000000   0.408402000000   1.143806000000
Be   0.905131000000   0.410037000000  -1.143131000000
B   -0.682576000000  -1.497315000000   0.425977000000
B   -0.684257000000   1.498360000000  -0.425282000000
B    0.684218000000  -1.498348000000  -0.426351000000
B    0.682942000000   1.498649000000   0.425686000000
B   -2.072794000000   0.728212000000  -0.251748000000
B    2.072272000000  -0.728536000000  -0.252462000000
B    2.071532000000   0.729531000000   0.251631000000
B   -2.071915000000  -0.730003000000   0.251800000000
12    
0.000000000         cluster_0001.out
Be  -0.905655000000  -0.409167000000   1.143981000000
Be   0.906419000000  -0.409960000000  -1.144242000000
Be  -0.905173000000   0.408402000000  -1.143806000000
Be   0.905131000000   0.410037000000   1.143131000000
B   -0.682576000000  -1.497315000000  -0.425977000000
B   -0.684257000000   1.498360000000   0.425282000000
B    0.684218000000  -1.498348000000   0.426351000000
B    0.682942000000   1.498649000000  -0.425686000000
B   -2.072794000000   0.728212000000   0.251748000000
B    2.072272000000  -0.728536000000   0.252462000000
B    2.071532000000   0.729531000000  -0.251631000000
B   -2.071915000000  -0.730003000000  -0.251800000000 
\end{verbatim}
\typeout{get arXiv to do 4 passes: Label(s) may have changed. Rerun}
\end{document}